\documentclass{aa}
\usepackage{graphicx}

\usepackage{wasysym}

\usepackage{longtable, lscape}

\newcommand{\Teff}{\mbox{$T_{\rm{eff}}$}}
\newcommand{\logg}{\mbox{log $g$}}
\newcommand{\feh}{\mbox{[Fe/H]}}

\newcommand{\cfe}{\mbox{[C/Fe]}}
\newcommand{\ofe}{\mbox{[O/Fe]}}
\newcommand{\oh}{\mbox{[O/H]}}
\newcommand{\co}{\mbox{[C/O]}}

\newcommand{\eg}{e.g.~\/}
\newcommand{\Ci}{C\,{\sc i}}
\newcommand{\Oi}{O\,{\sc i}}
\newcommand{\Fei}{Fe\,{\sc i}}
\newcommand{\Feii}{Fe\,{\sc ii}}

\begin{document}

  \title{
The C/O ratio at low metallicity:\\
constraints on early chemical evolution from observations of Galactic halo stars
\thanks{Based on data collected with the European Southern Observatory's {\it Very Large
Telescope} (VLT) at the Paranal, Chile (programmes No. 67.D-0106 and
73.D-0024) and with the {\it Magellan Telescope} at Las Campanas
Observatory, Chile}
        }

  \author{D. Fabbian\inst{1,2},
     P.~E. Nissen\inst{3},
     M. Asplund\inst{4},
     M. Pettini\inst{5},
     C. Akerman\inst{5}}

  \offprints{D. Fabbian, \\
\email{damian@mso.anu.edu.au} }

  \institute{Research School of Astronomy \& Astrophysics, The Australian
  National University, Mount Stromlo Observatory, Cotter Road, Weston
  ACT 2611, Australia
        \and Current address:  Instituto de Astrof\'isica de Canarias,
        Calle Via L\'actea s/n, E38205, La Laguna, Tenerife, Espa\~na
        \and Department of Physics and Astronomy, University of Aarhus,
        8000 Aarhus C, Denmark
        \and Max Planck Institute for Astrophysics, Postfach 1317, 85741
        Garching b. M\"unchen, Germany
        \and Institute of Astronomy, University of Cambridge,
        Madingley Road, Cambridge CB3 0HA, UK
}

  \date{Received / Accepted }

  \abstract
  {} 
  {We present new measurements of the abundances of carbon and oxygen
  derived from high-excitation \Ci\ and \Oi\ absorption lines in
  metal-poor halo stars, with the aim of clarifying the main sources of these
  two elements in the early stages of the chemical enrichment of the Galaxy.}
  {We target 15 new stars compared to our previous study, with an emphasis on
  additional C/O determinations in the crucial metallicity range
  $-3\apprle\feh\apprle-2$. The stellar effective temperatures were
  estimated from the profile of the H$\beta$ line. Departures
  from local thermodynamic equilibrium were accounted for in the
  line formation for both carbon and oxygen. The non--LTE effects are very strong at the
  lowest metallicities but, contrary to what has sometimes been assumed in the
  past due to a simplified assessment, of different degrees for the two elements. In
  addition, for the 28 stars with \feh\,$< -1$ previously analysed,
  stellar parameters were re-derived and non--LTE corrections
  applied in the same fashion as for the rest of our sample, giving consistent
  abundances for 43 halo stars in total.}
  {The new observations and non--LTE calculations strengthen previous
  suggestions of an upturn in C/O towards lower metallicity (particularly
  for \oh\,$\la -2$). The C/O values derived for these very metal-poor
  stars are, however, sensitive to excitation via the still poorly
  quantified inelastic H collisions. While these do not significantly
  affect the non--LTE results for \Ci, they greatly modify the \Oi\
  outcome. Adopting the H collisional cross-sections
  estimated from the classical Drawin formula leads to [C/O]$\approx
  0$ at [O/H]$\approx -3$. To remove the upturn in C/O,
  near-LTE formation for \Oi\ lines would be required, which could only
  happen if the H collisional efficiency with the Drawin recipe is
  underestimated by factors of up to several tens of times, which we
  consider unlikely.}
  {The high C/O values derived at the lowest metallicities may be revealing
  the fingerprints of Population III stars or may signal rotationally-aided
  nucleosynthesis in more normal Population II stars.}


  \authorrunning{D. Fabbian et al.}
  \titlerunning{The evolution of the C/O ratio in the Galactic halo}
  \maketitle


\section{Introduction}

Carbon and oxygen play a fundamental role in the chemical evolution
of the Universe. They rank as the most common elements produced via
stellar life and death, and their abundances are surpassed only by
those of H and He, which are instead linked to the Big Bang.

The abundances of carbon and oxygen may be important in relation to
a transition from massive Population III stars to a low-mass
Population II star formation mode. The latter encompasses the birth
mechanism of the oldest stellar population currently known,
including the most iron-deficient stars in the halo, which are
thought to carry the fingerprints of at most a few supernovae. The
very first stars in the Universe are predicted to have been very
massive, because the absence of metals (and thus, of cooling by
fine-structure lines of C and O) only allowed fragmentation on large
scales. The first lower-mass ($\apprle 1 M_{\odot}$) stars would
then only have formed once a critical metallicity (allowing cloud
fragmentation into smaller clumps) was reached in the early Universe
thanks to enrichment of elements ejected from the first supernovae.
Bromm \& Loeb (2003) suggest that no low-mass dwarf star should
exist having ({\it simultaneously}) [C/H]\,$\apprle -3.5$ {\it and}
\oh\,$\apprle -3$. They also point out that, in order to sample
individual supernova events occurring in the earliest epochs, the
best candidates among these second generation stars would be those
with abundances of carbon and oxygen very close to this critical
metallicity. Frebel, Johnson \& Bromm (2007) predict that all stars
with [Fe/H]\,$< -4$ should show enhanced C and/or O abundances,
because otherwise they would not have lived long enough (have low
enough mass) to be observed.

In this context, it is interesting to note that, very recently,
Carollo et al. (2007) have highlighted that our Galaxy has a second,
more distant halo structure (the ``outer halo'') with a lower peak
metallicity and probably different (dissipationless) formation
mechanism than the inner halo.

%
Despite our knowledge of the nuclear processes involved (see e.g.
reviews by Wallerstein et al. 1997 and El Eid 2005), the constraints on
the actual sources of carbon in the Galaxy are still not satisfying.
There is ongoing debate on whether the bulk of carbon yields is
mainly contributed by massive stars (\eg Carigi et al. 2005) or low-
and intermediate-mass stars (\eg Chiappini, Romano \& Matteucci
2003).

In a number of investigations (\eg Andersson \& Edvardsson 1994;
Gustafsson et al. 1999; Reddy, Lambert \& Allende Prieto 2006), the
[C/Fe] abundance ratio in the thin disc has been found to slowly
decrease with time and increasing metallicity. The work of  Bensby
\& Feltzing (2006) suggests that [C/Fe] flattens to roughly the
solar value at intermediately-low metallicities ($-1 <$[Fe/H]\,$<
0$). This trend is robust to non--LTE effects (Fabbian et al. 2006),
since that study employs the forbidden [\Ci] line at $8727$~\AA.
However, when moving to the halo stellar population, this absorption
feature becomes too weak. Akerman et al. (2004) have investigated
the derivation of accurate abundances from high-excitation infrared
\Ci\  lines, detected down to \feh\,$\sim -3.2$. Fabbian et al.
(2006) have pointed out how at these low metallicities, after
accounting for non--LTE effects, a roughly flat plateau is evident
at a level of [C/Fe]\,$\approx 0$, even though a relatively large
(and possibly real) scatter remains.

%
Oxygen is thought to be synthesized in short-lived massive stars,
which end their lives as type II SNe, dispersing their chemical
make-up into the interstellar medium (ISM). Despite many studies,
the chemical evolution of oxygen during the early history of the
Milky Way is still debated and not well understood, giving rise to
the so-called ``oxygen problem''. Different abundance indicators give
conflicting results for halo stars, either a linear increase with
decreasing metallicity, reaching [O/Fe]\,$\approx +1.0$ dex at \feh$=
-3$, when using UV OH lines (Israelian et al. 1998; Boesgaard et al.
1999) or a flat plateau when using the forbidden [\Oi] $6300$~\AA\,
line (Barbuy 1988; Nissen et al. 2002; Garc\'{\i}a P{\'e}rez et al.
2006), while the \Oi\ $7772-7775$~\AA\, triplet in metal-poor
unevolved stars typically implies values between those two extreme
trends. This problem is crucial, since the adopted oxygen abundance
influences, for example, the derived ages of globular clusters and
the production of Li, Be, and B from spallation of C, N, and O atoms
in the early Galaxy.

For oxygen there are several potential pitfalls in the analysis that
can result in systematic errors. The unresolved issue of the correct
effective temperature
($T_{\rm eff}$) scale at low metallicities
is clearly important, since it affects OH
and \Oi\ in opposite ways (Mel\'{e}ndez et al. 2006). For molecules
like OH, the very different atmospheric temperature structure in 3D
hydrodynamical model atmospheres compared with standard 1D
hydrostatic models leads to very large negative 3D abundance
corrections (Asplund \& Garc\'{\i}a P{\'e}rez 2001; Collet, Asplund
\& Trampedach 2007). The [\Oi] line is also sensitive to such 3D
effects but not as severely (Nissen et al. 2002).  While it has been
known for a long time that the \Oi\ $7772-7775$~\AA\, lines are
prone to departures from LTE (e.g. Kiselman 1991; Asplund et al.
2004), Fabbian et al. (2008a) have very recently demonstrated how
the relevant non--LTE corrections are likely to have been underestimated
due to inadequate collisional data being used in the construction of
the atomic models employed. The outcome of using an up-to-date such
model, including accurate electron collisional cross-sections
obtained through quantum mechanical calculations, is a sharp increase
in the non--LTE effects for \feh\,$\apprle-2.5$. The physical
explanation is in terms of radiative pumping in the UV resonance
lines and efficient intersystem collisional coupling.

%
Early observational results (Tomkin et al. 1992) showed that
[C/O] is subsolar when [O/H]\,$< 0$, and suggested that the
ratio remains essentially flat at this level down to low metallicities.
More recently, renewed attention has been given to investigating
the behaviour of  the \co\ ratio with decreasing metallicity.
By looking respectively at metal-poor dwarf stars
and giant stars in the halo of our
Galaxy, Akerman et al. (2004) and Spite et al. (2005)
suggested that \co\ increases again at very low metallicities,
recovering near-solar values when [O/H]\,$\simeq -3$.
Unfortunately, different abundance indicators were used in the
two investigations, so that the results of these two studies
may not be directly comparable.
However, the suggestion of high [C/O]
values at the lowest metallicities
is potentially very important and deserving of further
scrutiny, as it may be an indication of
C-rich ejecta from massive Population III SNe.

As reliable non--LTE corrections
to the C and O abundances
in late-type stars were not available at the
time, Akerman et al. (2004)
assumed that the \Ci\ lines near $9100$~\AA\, would be subject to
the same non--LTE effects as the IR oxygen triplet lines, given
that all these spectral features arise from highly excited levels.
However, recent analyses of the problem (Fabbian et
al. 2006, 2008a) have shown that the abundance corrections for both
elements are likely to be more negative than previously assumed. In
particular, for typical low-metallicity halo stars
such corrections amount to
$\sim -0.4$ dex for carbon and $\sim -0.9$ for oxygen,
ignoring the still very uncertain effects of inelastic H collision.
Including the collisions in accordance with the classical
Drawin (1968, 1969) recipe, reduces the magnitude of the
corrections which, however, remain significant
(in particular, $\sim -0.5$ dex for O).

%
Complementary information on the nucleosynthetic origin of C and O
at low metallicity is also available from high-redshift observations
of metal-poor damped Lyman-alpha systems (DLAs)
which are generally interpreted as galaxies in early stages
of chemical evolution (Wolfe,  Gawiser, \& Prochaska 2005).
For example, Erni et al. (2006) found the chemical abundances
in a DLA with \oh\,$\sim-2.5$ to be consistent with
enrichment from a single starburst of massive ($10-50\,M_{\odot}$)
zero-metallicity stars. Pettini et al. (2008) very recently
derived near-solar \co\, ratios in a small sample of
the most metal-poor DLAs/subDLAs known.

Becker et al. (2006) studied absorption toward high-redshift
($4.9 < z <6.4$) quasars, inferring a mean [C/O]\,$\sim -0.3$ in the
Lyman alpha forest clouds which trace the low-density
and low-metallicity intergalactic medium (IGM).
At lower redshifts, ($2.1 < z < 3.6$), Aguirre et al. (2008)
deduced [C/O]\,$= -0.66 \pm 0.2$.
While broadly in line with halo star abundances,
these IGM values are difficult to interpret because:
(a) they rely on the accuracy of large photoionisation corrections,
and (b) the origin of the metals found in the IGM is still
unclear (e.g. Ryan-Weber et al. 2006 and references therein).

In the present study, we aim to use halo stellar abundances as tracers of early
Galactic chemical evolution, by deriving carbon and oxygen
compositions in non--LTE for a new set of metal-poor stars, as
well as carrying out an improved analysis of previous data. The
results provide an important test for Galactic chemical evolution
models.

\section{Observations and data reduction}
\label{obs}

\begin{flushleft}

\begin{table}
\begin{center}
\begin{minipage}{0.4\textwidth}
\caption{The IDs and atmospheric parameters (effective temperature,
gravity and iron content) of the halo stars observed.
Our sample includes 28 stars from the 2001 UVES run and having \feh\,$< -1$.
For such objects \Teff\ was re-determined by us consistently with
the rest of the sample.
\label{fabt:parameters}}
\begin{footnotesize}
\begin{tabular}{p{2.5cm}ccc}
\hline \hline
ID & \Teff & \logg & \feh\\
\hline
{\bf UVES (2001)} & & &\\
BD-13$^{\circ}$3442  & 6366 & 3.99 & -2.69\\
CD-30$^{\circ}$18140 & 6272 & 4.12 & -1.89\\
CD-35$^{\circ}$14849 & 6294 & 4.26 & -2.34\\
CD-42$^{\circ}$14278 & 6085 & 4.39 & -2.03\\
G011-044             & 6178 & 4.35 & -2.03\\
G013-009             & 6343 & 4.01 & -2.29\\
G018-039             & 6093 & 4.19 & -1.46\\
G020-008             & 6194 & 4.29 & -2.19\\
G024-003             & 6084 & 4.23 & -1.62\\
G029-023             & 6194 & 4.04 & -1.69\\
G053-041             & 5993 & 4.22 & -1.29\\
G064-012             & 6435 & 4.26 & -3.24\\
G064-037             & 6432 & 4.24 & -3.08\\
G066-030$^*$         & 6470 & 4.29 & -1.48\\
G126-062$^*$         & 6224 & 4.11 & -1.55\\
G186-026             & 6417 & 4.42 & -2.54\\
HD106038             & 6027 & 4.36 & -1.37\\
HD108177             & 6156 & 4.28 & -1.71\\
HD110621             & 6157 & 4.08 & -1.59\\
HD140283             & 5849 & 3.72 & -2.38\\
HD160617             & 6047 & 3.84 & -1.75\\
HD179626             & 5881 & 4.02 & -1.12\\
HD181743             & 6044 & 4.39 & -1.87\\
HD188031             & 6234 & 4.16 & -1.72\\
HD193901             & 5699 & 4.42 & -1.10\\
HD194598             & 6020 & 4.30 & -1.15\\
HD215801             & 6071 & 3.83 & -2.28\\
LP815-43             & 6483 & 4.21 & -2.71\\
\hline
{\bf UVES (2004)} & & &\\
CD-24~17504          & 6338 & 4.32 & -3.21\\
CD-71~1234$^*$       & 6325 & 4.18 & -2.38\\
CS 22943-0095        & 6349 & 4.18 & -2.24\\
G004-037             & 6308 & 4.25 & -2.45\\
G048-029$^{**}$        & 6482 & 4.25 & -2.60\\
G059-027$^*$         & 6272 & 4.23 & -1.93\\
G126-052             & 6396 & 4.20 & -2.21\\
G166-054             & 6407 & 4.28 & -2.58\\
HD84937              & 6357 & 4.07 & -2.11\\
HD338529             & 6373 & 4.03 & -2.26\\
LP635-014            & 6367 & 4.11 & -2.39\\
LP651-004            & 6371 & 4.20 & -2.63\\
\hline
{\bf MIKE (2003)} & & &\\
G041-041             & 6440 & 4.06 & -2.66\\
G048-029$^{**}$      & 6489 & 4.25 & -2.61\\
G084-029             & 6302 & 4.05 & -2.62\\
LP831-70             & 6232 & 4.36 & -2.93\\
\hline
                        & & &\\

\end{tabular}
\end{footnotesize}

$^*$Single-lined spectroscopic binary stars.\\


$^{**}$This star was observed in both the UVES (2004) and
MIKE (2003) runs.

\end{minipage}
\end{center}
\end{table}

\end{flushleft}


\begin{flushleft}

\begin{figure*}[ht]
  \centering
  \includegraphics[width=5.9cm]{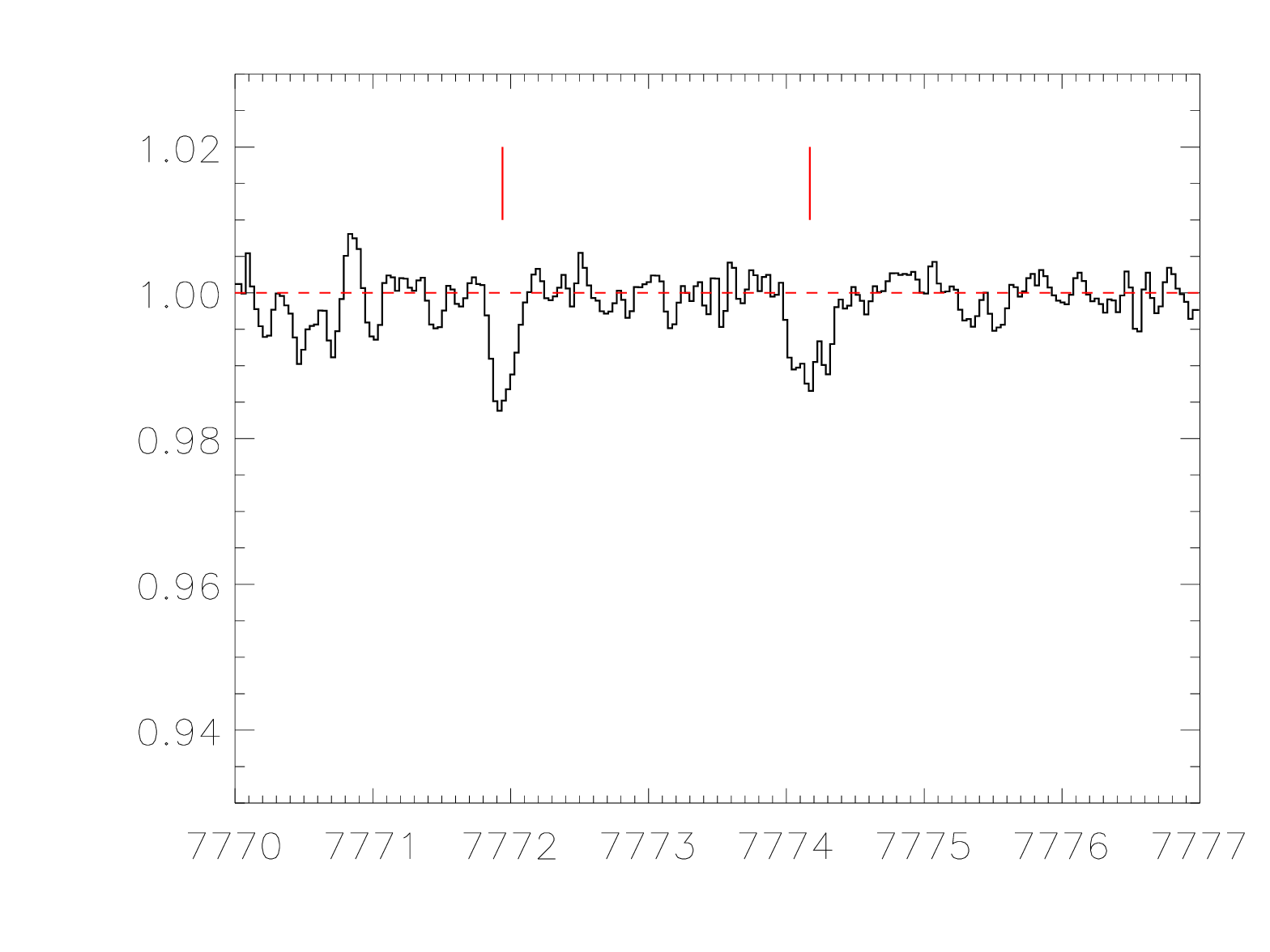}
  \includegraphics[width=5.9cm]{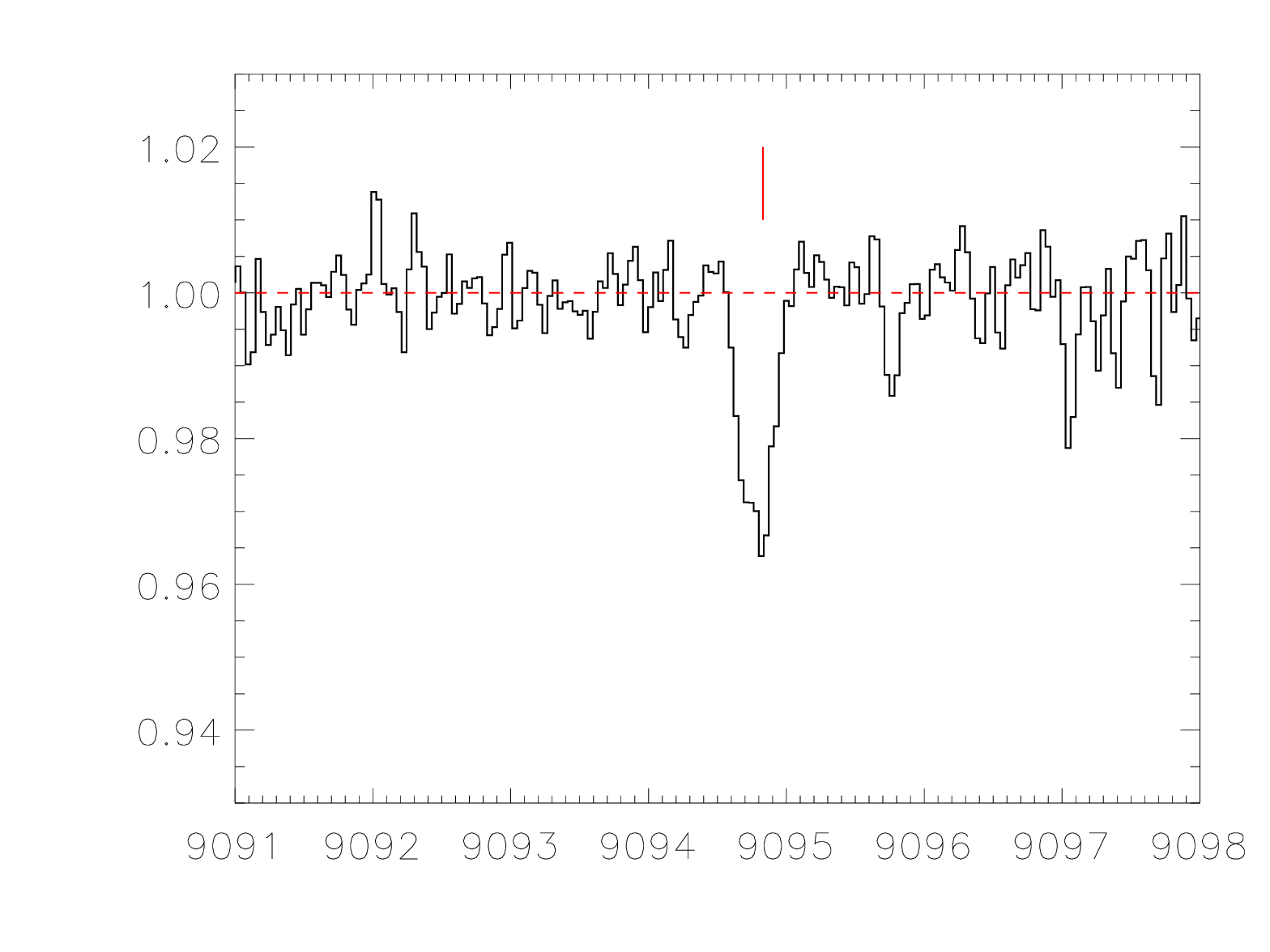}
  \includegraphics[width=5.9cm]{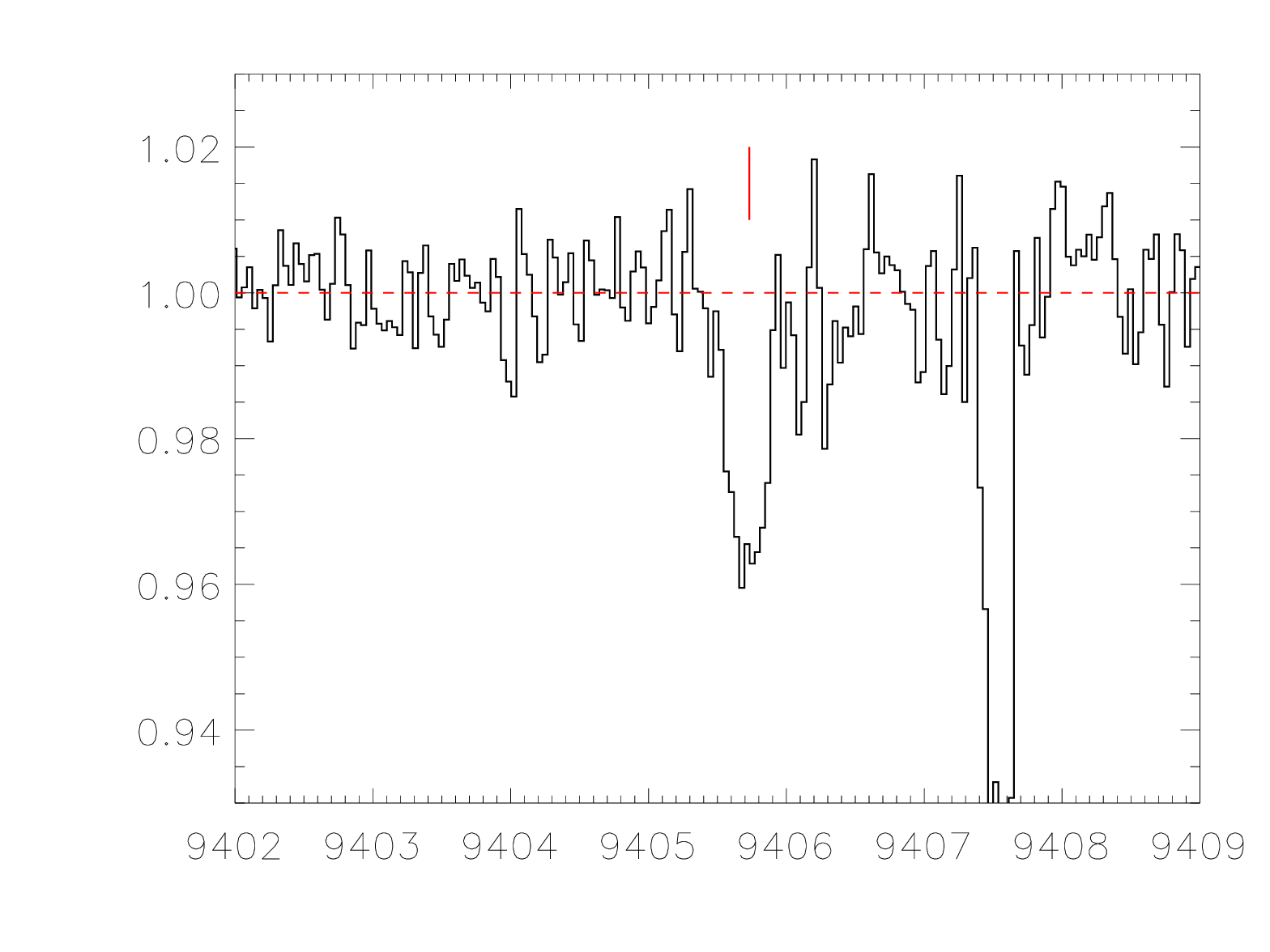}
  \includegraphics[width=5.9cm]{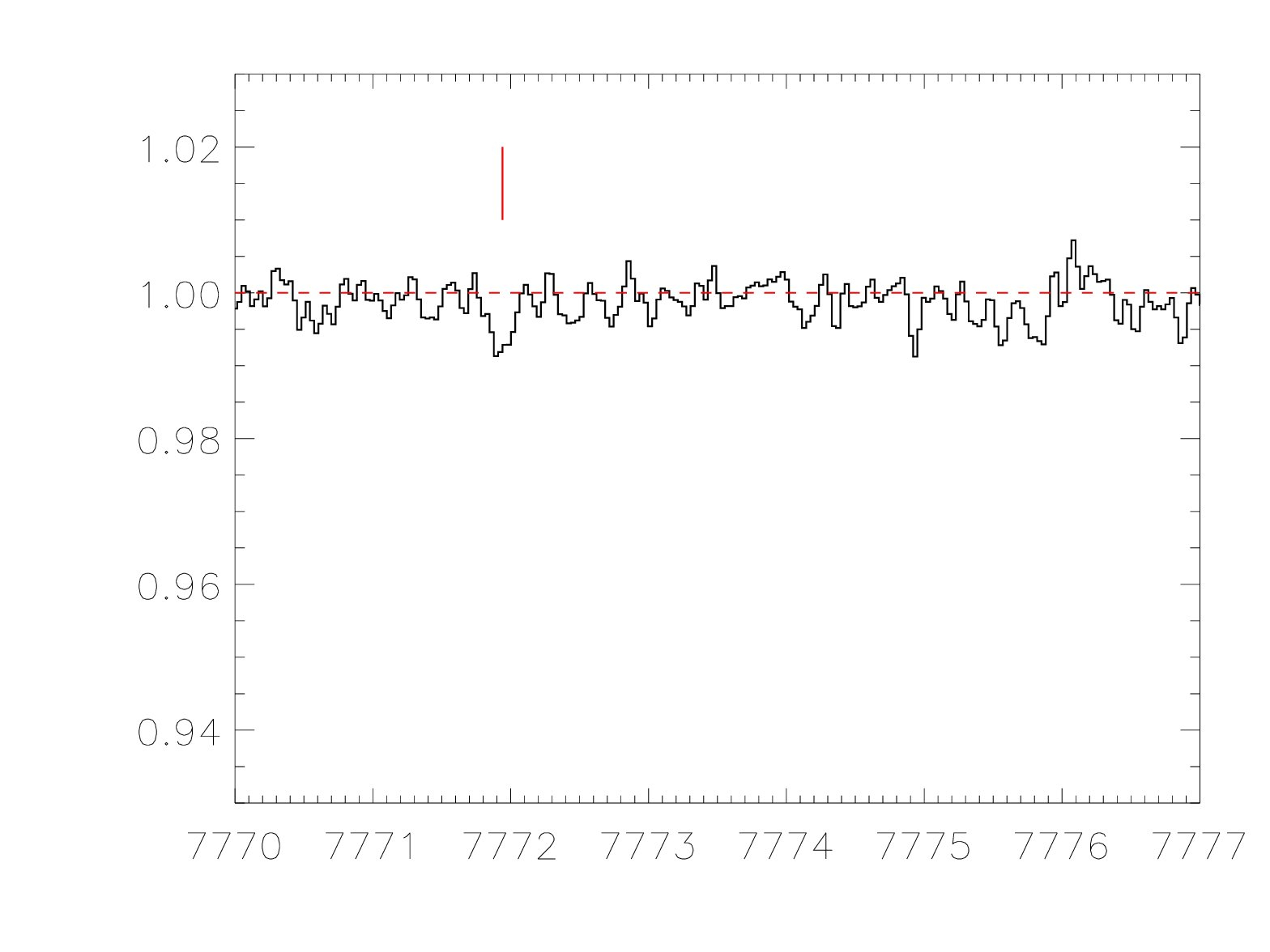}
  \includegraphics[width=5.9cm]{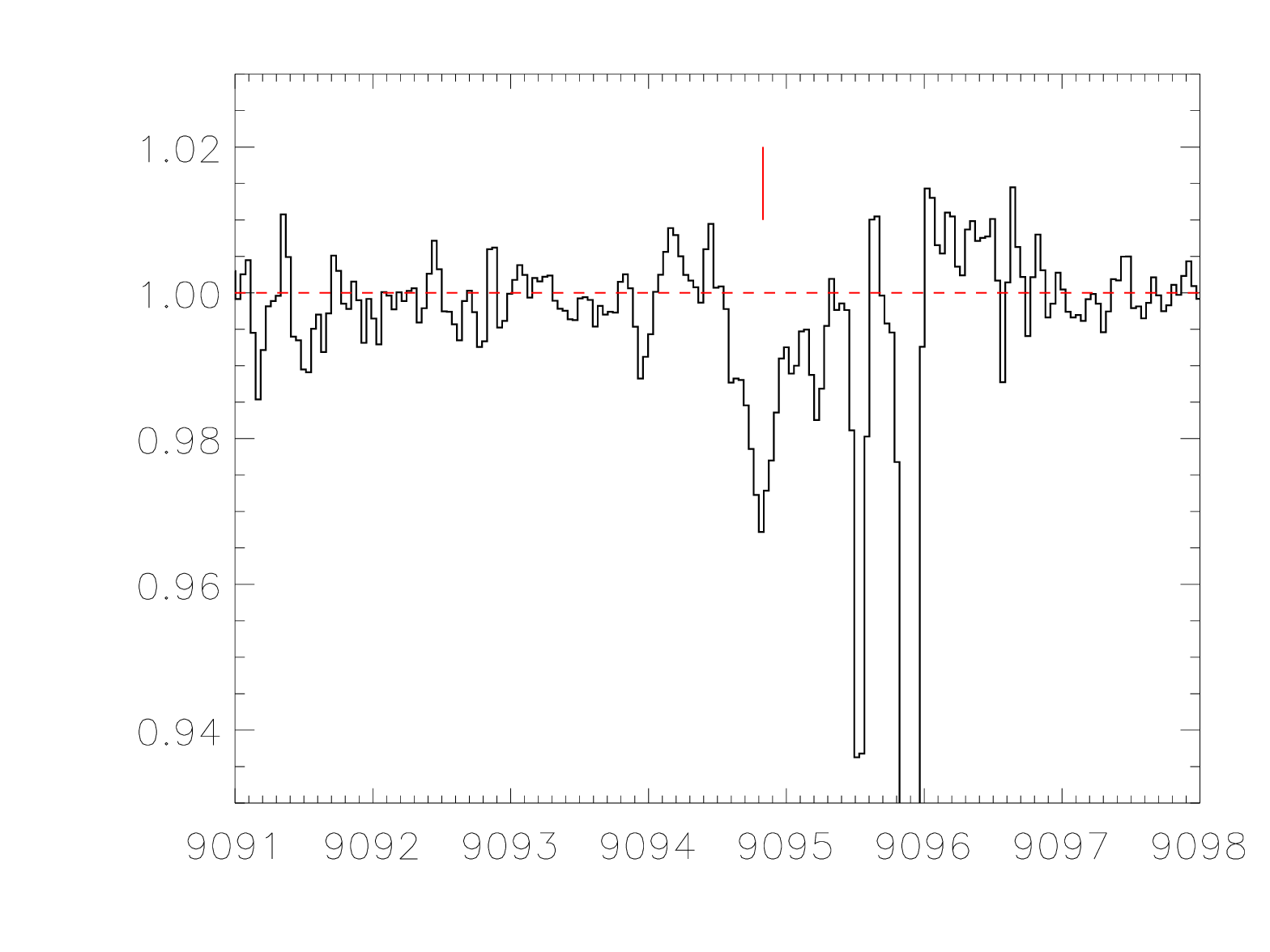}
  \includegraphics[width=5.9cm]{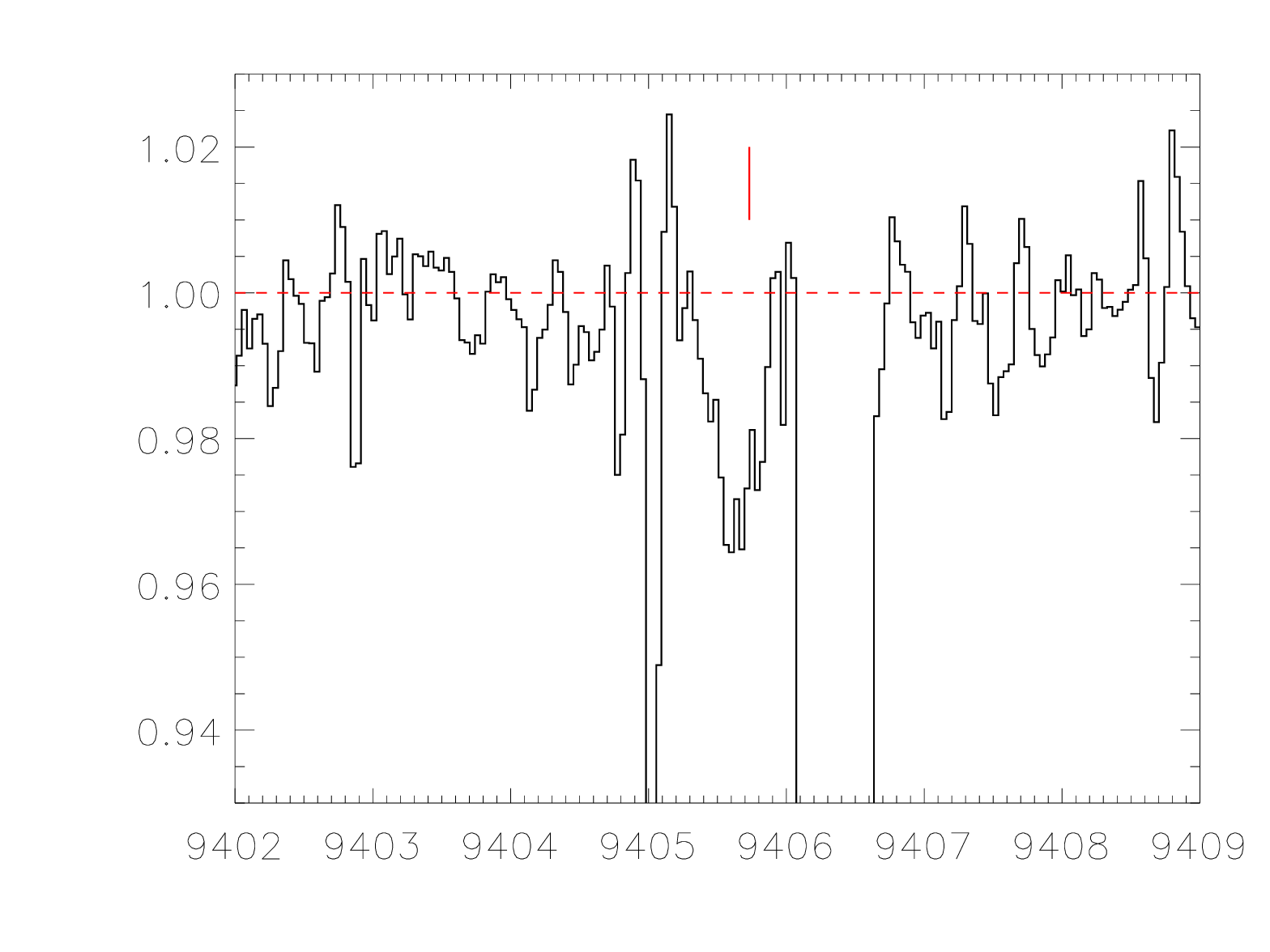}\\
  \caption{The spectral quality in the regions near the \Oi\
  (left panels) and \Ci\ (middle and right panels)
  high-excitation lines is shown for two of the most metal-poor
  stars (G64-37, at [Fe/H]$=-3.08$, upper
  panels, and CD-24~17504, at [Fe/H]$=-3.21$, lower panels) in the sample,
  with wavelength (in \AA) on the horizontal axis and
  flux (normalized to the continuum level) on the vertical one.
  The spectra have S/N$\sim 200$ and $\sim 400$, respectively,
  in the \Oi\, triplet region. The S/N around the
  \Ci\, lines is a factor $\sim 2$ lower, and strong residuals from the
  removal of telluric lines are present. Spectral features of interest near
  $7774$, $9095$, and $9406$ \AA\, for
  which we could determine reliable equivalent width estimates in these
  two stars are marked
  by short vertical ticks above the observed
  spectra.
  \label{fabf:spectra}}
\end{figure*}

\end{flushleft}


The stellar sample employed in this study is mainly composed of
observations carried out at the European Southern Observatory's Very
Large Telescope (ESO VLT) using its two-arm, cross-dispersed,
high-resolution Ultraviolet-Visual Echelle Spectrograph (UVES, see
Dekker et al. 2000) mounted on the UT2 (Kueyen telescope).

The dataset is composed of spectra for 12 very metal-poor stars
observed in 2004, together with spectra of 28 halo stars with
\feh$<-1$ (from Akerman et al. 2004) for which we
re-derived stellar parameters in a consistent fashion with the new
stars. This UVES sample is the same as that used by Nissen et al. (2007);
therefore we direct the reader to the description of the
observations therein. Here we just mention that the UVES spectra
cover the blue region $3800-5000$~\AA\ and the near-IR region
$6700-10500$~\AA, both with a resolving power
$R \equiv \lambda / \Delta \lambda
\simeq 60\,000$. The blue region contains \Fei\ and \Feii\ lines
from which we determine [Fe/H], and the H$\beta$ line used for deriving
the effective temperature of the stars. The \Ci\, and \Oi\ lines are
found in the near-IR region.

Spectra of four additional stars were obtained with the Magellan
6.5~m telescope over three nights in December 2003. The Magellan
Inamori Kyocera Echelle (MIKE), a high-throughput double echelle
spectrograph (Bernstein et al. 2003), was used. The entrance slit
was set at $0.35$ arcsec, which corresponds to a resolving power $R
\simeq 60\,000$, i.e. the same as the UVES spectra, but with only
two pixels per spectral resolution element compared to four pixels
for the near-IR UVES spectra. The MIKE spectra cover, however,
broader spectral ranges than UVES, i.e. $3300-4900$~\AA\, in the
blue arm and $4900-10000$~\AA\, in the red arm. Actually, the blue
part was not used, mainly because the H$\beta$ line is not
sufficiently well centered in an echelle order to derive $T_{\rm
eff}$ with the technique applied by Nissen et al. (2007). Instead,
$T_{\rm eff}$ was derived from H$\alpha$ as described by Asplund et
al. (2006), and [Fe/H] was determined from \Feii\, lines in the
$4900-6700$~\AA\, region.

Since one of the stars in the MIKE sample, G48-29, was also observed
in our UVES 2004 run, this brings the total of newly observed
targets to 15. Together with the observations by Akerman et al.
(2004), a total of 43 halo stars are thus employed in the present
study.

The new spectra were reduced with the standard echelle data
reduction package in IRAF. Background subtraction, flat-fielding,
order extraction and wavelength calibration were done using
semi-automatic IRAF routines. The \Ci\, lines are located in a part
of the optical spectrum affected by the presence of strong telluric
water vapour lines. Thus, spectra were obtained and processed not only
for the programme stars but also for hot, fast-rotating, B-type
stars, which are necessary to effectively correct for such
atmospheric disturbance in the infrared. The IRAF task ``telluric''
was used to divide the spectrum of our sample stars by those of the
comparison B-type stars. We then used the task ``continuum'' to
normalize the spectra (continuum placement). We corrected for the
radial velocity Doppler shift by measuring the wavelengths of a few
selected lines. Finally, equivalent widths were measured on the
reduced spectra for the lines of interest by performing a Gaussian
fit, or by direct integration in the case of very weak lines or poor
fit due to noisy spectral regions.

The reduction of the MIKE spectra was complicated by the fact that
the image of the slit is significantly tilted and curved on the CCD.
Nevertheless, once this was accounted for, the final quality of the
data proved to be similar to that of the UVES spectra.
For stars with repeated observations on different nights
we co-added the individual spectra to improve the final
signal-to-noise ratio. Overall, the new data are of
similar, or higher, quality than those in the study
by Akerman et al. (2004).
The typical signal-to-noise ratios per pixel around the \Ci\, and \Oi\,
spectral features of interest are S/N\,$=200-400$, the
region encompassing the oxygen triplet having higher S/N than that around
the high-excitation carbon lines (see Fig.\,\ref{fabf:spectra}).

Generally, all spectral lines of interest in this study were
detectable in the UVES spectra thanks to the relatively high S/N
obtained. Even in the case of \object{CD-24~17504}, the most metal-poor (\feh\,$=-3.21$)
star among the 15 newly observed,
we were able to derive reliable estimates of the
carbon and oxygen abundances from the strongest lines in the
multiplets targeted.
For oxygen in particular,
the equivalent width measured is only $1.7$\,m\AA, but the
reliability of the derived abundance was confirmed via spectral
synthesis (see Subsection \ref{fabss:LTE}). Regarding the Magellan
observations, it was possible to achieve on average similarly high
S/N. This is somehow offset by lower pixel sampling than in the UVES
spectra and by rapid drop in the MIKE CCD sensitivity beyond $9200$\,\AA,
the latter implying that the S/N is too low for a detection at low
metallicity of \Ci\, $9405.7$\,\AA\, (which also falls very close to
the edge of an echelle order) in the midst of strong telluric lines.
This therefore makes \Ci\ $9094.8$\,\AA\, and $9111.8$\,\AA\, the
only useful lines for deriving the carbon abundance in this case.
These \Ci\ lines are however not clearly detectable in \object{LP831-70},
the most metal-poor object in the sample observed with MIKE, even
though the star was observed on all three nights and all spectra
were combined to achieve a high S/N\,$\sim 380$ per pixel. In
particular, due to an overlapping water vapour feature, even the
stronger $9094.8$\,\AA\, line is not measurable.
We estimated upper limits to the carbon and oxygen abundances in this star
by measuring the equivalent widths (typically a few m\AA) of noise
features.

\section{Elemental abundance analysis}

\subsection{Stellar parameters}
\label{fabss:stparam}

Estimates of the effective temperatures for all stars in this study,
including those with \feh\,$<-1$ in the Akerman et al. (2004) sample,
were derived using the H$\beta$ line profile. For a detailed
discussion of the procedure and the improved accuracy of the
\Teff\ determinations, see Nissen et al. (2007). As discussed
therein, differential values are determined with a {\it precision}
of $\sim 30$~K for metal-poor turnoff stars. Below, we compare with
effective temperature estimates from other methods.

\begin{flushleft}

\begin{figure}
  \includegraphics[width=8.9cm]{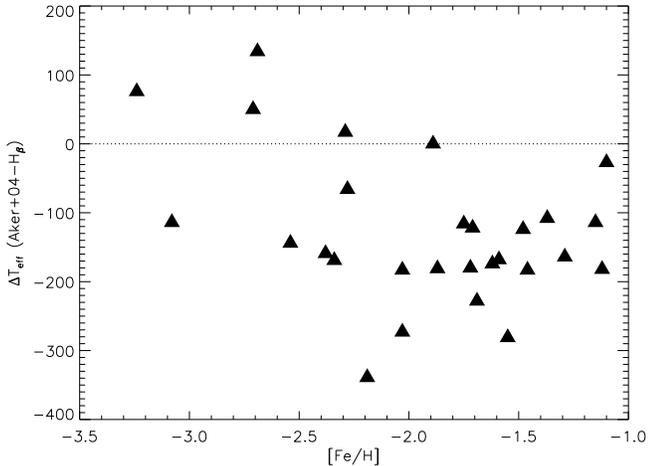}
  \caption{
  Run of $\Delta$\Teff\, (the difference between effective temperature
  determinations in Akerman et al. 2004 and in this work) against
  metallicity.
\label{fabf:delta_teffs}}
\end{figure}

\end{flushleft}

Akerman et al. (2004) derived their \Teff\, values from ($b\!-\!y)$ and
($V\!-\!K)$ colour indices. Fig.~\ref{fabf:delta_teffs} shows the
difference between their and our \Teff\, values for the stars
with \feh\,$ < -1$ in
common between the two studies.
For the six metal-rich stars with \feh\,$>-1$
in the Akerman et al. sample,
it is very hard to determine accurate temperatures
from H$\beta$ because of many blending metal lines across
its profile. Those stars are not included in the present investigation,
since we are mainly interested in the behaviour of C/O at
very low \feh.
Inspection of Fig.~\ref{fabf:delta_teffs}
shows that our temperature estimates are typically $150$\,K
higher than those by Akerman et al. (2004) down to \feh\,$\sim -2.6$, with a
few stars showing a difference of up to $\sim 300$\,K, while a few
others have very similar determinations in the two studies. In
contrast, our estimated effective temperatures tend to be lower (for
three out of four stars) at very low metallicities.


The advantage of the H$\beta$ method is that errors in
gravity, metallicity and in interstellar reddening do not affect
the determination of \Teff.
A comparison with \Teff\ values based on $V-K$
calibrations (Alonso et al. 1996; Ram{\'\i}rez \& Mel\'{e}ndez
2005b; Masana, Jordi \& Ribas 2006) shows that the difference with
those estimates tends to switch sign (and become larger) with
decreasing metallicity. The H$\beta$-based temperatures are higher
by 50--100\,K when \feh\,$> -2$, but lower by about 100\,K
for \feh\,$<-2.5$.
In the transition region, $-2.5 \le\feh\le -2$, there is a
large scatter (Nissen et al. 2007). This is essentially the same
behaviour as just discussed when comparing with the
estimates by Akerman et al. (2004),
except that in that case the residuals are larger in the
\feh\,$> -2$ regime.
This is due to a systematic offset
of $\sim 50$--100\,K in the values of \Teff\
derived by Akerman et al. (2004)
compared with those by Ram{\'\i}rez \& Mel\'{e}ndez (2005b) and
Masana et al. (2006).

Masana and co-workers suggested that temperature estimates
from the infrared flux method
(IRFM) may be too high by $\sim 200$\,K for
\feh$\,< -2.5$.
However, the $V-K$ calibrations by
Masana et al. are in fairly good agreement with the
Ram{\'\i}rez \& Mel\'{e}ndez scale for our sample of stars.
On the other hand, in their calibrations, Masana et al.
give two equations: one valid for $0.35 < (V-K)_0 < 1.15$ and
the other for $1.15 \apprle (V-K)_0 < 3.0$.
One would then
expect a continuous transition between \Teff\ estimates obtained with
the two calibrations.
However, this is not the case.
Since many
metal-poor turnoff stars have indeed $(V-K)_0$ values close to
1.15, the final result will depend on whether a star, after
reddening correction, happens to fall below or above that value,
with important differences in the \Teff\, derived in the two cases,
amounting to $\sim 200$\,K at very low \feh. In particular, most of
the largest discrepancies ($>100$\,K) occur for values around the
mentioned discontinuity, within $\pm 0.1$ of $(V-K)_0 = 1.15$.
It therefore appears that the Masana et al. calibrations may
systematically overestimate \Teff\, around $(V-K)\approx 1.15$.

In any case, this apparent inconsistency can not straightforwardly
explain the differences with the H$\beta$-derived temperatures,
since the resulting \Teff\  values from the equations by Masana et
al. agree reasonably well with those from the Ram{\'\i}rez \&
Mel\'{e}ndez calibrations, which apparently have no such
inconsistencies. Our H$\beta$-derived temperatures may thus be too
low by $\sim 100$\,K at the lowest metallicities. In general,
$(V-K)$ calibrations are bound to be less effective at very low
metallicities, because of the small numbers of such stars.
Furthermore, these objects are affected by an uncertain degree of
reddening, because they tend to be fainter and more distant.
Finally, the $(V-K)$ colour tends to saturate in metal-poor turnoff
stars and is, hence, less sensitive to \Teff\  (see Fig. 9 of
Ram{\'\i}rez \& Mel\'{e}ndez 2005a). The discrepancy between \Teff\
determinations derived with the various methods does indicate that
the effective temperature scale for metal-poor stars is still
uncertain, and that a "hotter" temperature scale at low \feh\ is not
warranted. It is clear that further improvements in model
atmospheres and line broadening theory, consideration of possible
non--LTE effects on Balmer lines, and other factors will need to be
explored in order to obtain fully consistent results. We estimate
our temperatures to be determined within $\pm 100$\,K in an {\it
absolute} sense. This uncertainty has a significant repercussion on
the determination of the absolute carbon and oxygen abundances, and
it is therefore important in the context of the interpretation of
element ratios which are relevant to Galactic chemical evolution,
such as  C/Fe and O/Fe. On the other hand, the C/O ratio is hardly
affected by systematic uncertainties in the \Teff\ scale, since the
high-excitation \Ci\ and \Oi\ lines we employ show similar
dependences on \Teff. A change in \Teff\ would affect the C and O
abundances by comparable amounts, thereby preserving our final C/O
estimates.

\begin{flushleft}

\begin{figure}
  \centering
  \includegraphics[width=8.9cm]{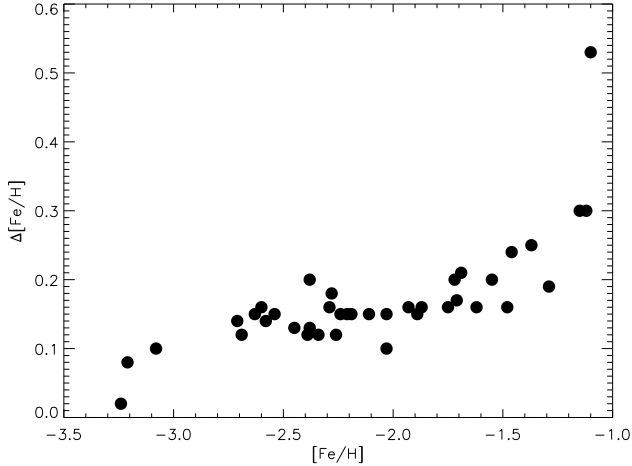}
  \caption{$\Delta {\rm [Fe/H]} = {\rm [Fe/H]}_{\rm II} -  {\rm [Fe/H]}_{\rm I}$
   (i.e., mean difference in the abundance of Fe as deduced
   from \Feii\, and \Fei\, lines) vs. [Fe/H]$_{\rm II}$,
  for the stars with determinations from both ionization stages.
   \label{fabf:Fe}
   }
\end{figure}

\end{flushleft}


The derivation of the other atmospheric parameters (surface gravity,
metallicity and microturbulence estimates) was also as described by
Nissen et al. (2007). All atmospheric parameters were determined by
iterating until consistency was achieved, with final adopted values
as listed in Table \ref{fabt:parameters}. Gravities are derived from
both $uvby$-$\beta$ photometry and Hipparcos parallaxes, except in
one case where neither was available.\footnote{CD~$-71\hbox {$^\circ
$ }1234$, for which we derived \logg\ by matching the difference
between Fe\,{\sc i} and Fe\,{\sc ii} abundances in the star with the
average difference between the two sets of Fe abundances in the
other stars.} Iron abundances were derived from \Feii\ lines, to
take advantage of the relative insensitivity of such features to
non--LTE effects (Asplund 2005). The statistical errors on the \feh\
values are of the order of $0.05-0.10$\,dex. A number of unblended
\Fei\ lines of similar strength as the \Feii\ lines were also
measured in the UVES spectra, to check on the \Fei/\Feii\ ionisation
equilibrium and on possible indications that non--LTE effects may
have been underestimated. By comparing average abundances derived
from lines in the two ionisation stages, we found that the residuals
have a tendency to decrease with decreasing metallicity (see
Fig.\,\ref{fabf:Fe}). The expectation is that non--LTE effects work
in the sense of producing lower Fe abundance estimates from \Fei\
lines compared with \Feii\ (Asplund 2005). In our case, the average
difference amounts to $\sim 0.15$\,dex. The trend we find would
point to differential, and metallicity-dependent, non--LTE effects
for \Fei\ relative to the Sun, or to adopted values of \Teff\ which
are too low (or possibly to a combination of both). The most
metal-rich star in our sample (\object{HD~193901}) is the only one
exhibiting a large disagreement between \Fei- and \Feii- based Fe
abundances. This may be due to the fact that the \Fei\ abundance is
based on only one line in HD~193901. Moreover, this line being quite
strong, the abundance derived from it depends critically on the
assumed microturbulence and damping constant. The [Fe/H] values in
Table \ref{fabt:parameters} are based on the \Feii\ lines, relative
to an adopted solar iron abundance of $\log \epsilon ({\rm
Fe})_{\odot } = 7.45$ (Asplund, Grevesse, \& Sauval 2005).

As mentioned in Sect. 2, \Teff\ for the MIKE stars was determined
from H$\alpha$ instead of H$\beta$. Taking into account that Nissen
et al. (2007) found a systematic difference of 64\,K between the
\Teff\ scale based on H$\beta$ and the one based on H$\alpha$, we
have thus transformed from the \Teff(H$\alpha$) scale to the
\Teff(H$\beta$) scale accordingly. The other model atmosphere
parameters were consistently derived as follows: (a) surface
gravities from Stromgren photometry and Hipparcos parallax (only
available for \object{G84-29}); (b) [Fe/H] from the same \Feii\
lines as in Asplund et al. (2006).

The increase of \Teff\ by $64$~K leads to a change of [Fe/H] by
$+0.01$ to $+0.02$\,dex only. The corresponding change of [C/H] and
[O/H] is $-0.03$\,dex.  Our choice of using H$\alpha$ in the
determination of \Teff\ for the MIKE spectra does not affect the
deduced values of C/O and thus will not impact our final results (see
also Fabbian et al. 2006). Note that in the case of \object{G48-29}, which was
observed with both MIKE and UVES, there is good agreement between the
atmospheric parameters estimated from the two sets of data (Table
\ref{fabt:parameters}).

\subsection{LTE abundance analysis}
\label{fabss:LTE}


\begin{flushleft}

\begin{figure*}[ht]
  \centering
  \includegraphics[width=8.9cm]{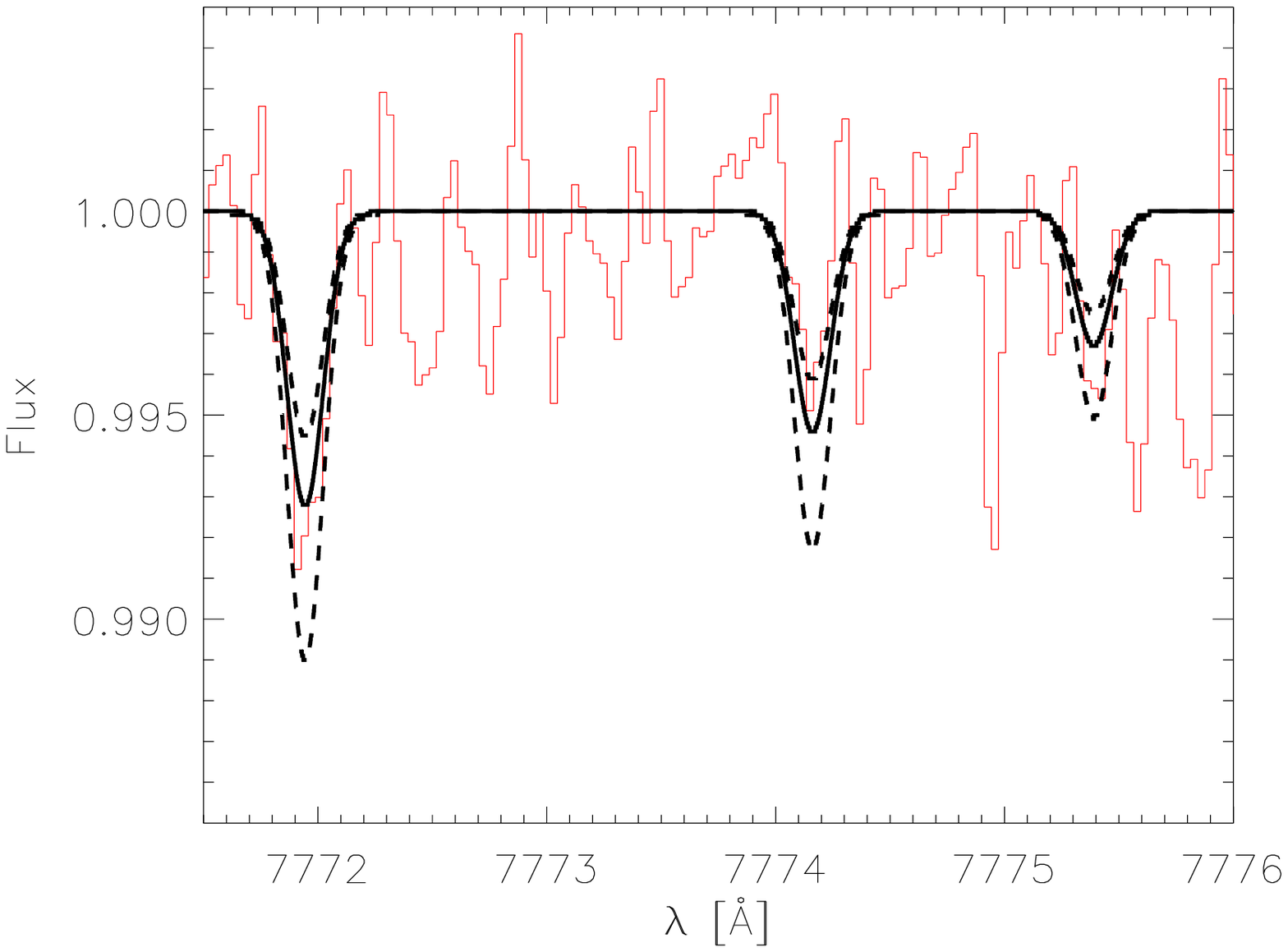}
  \includegraphics[width=8.9cm]{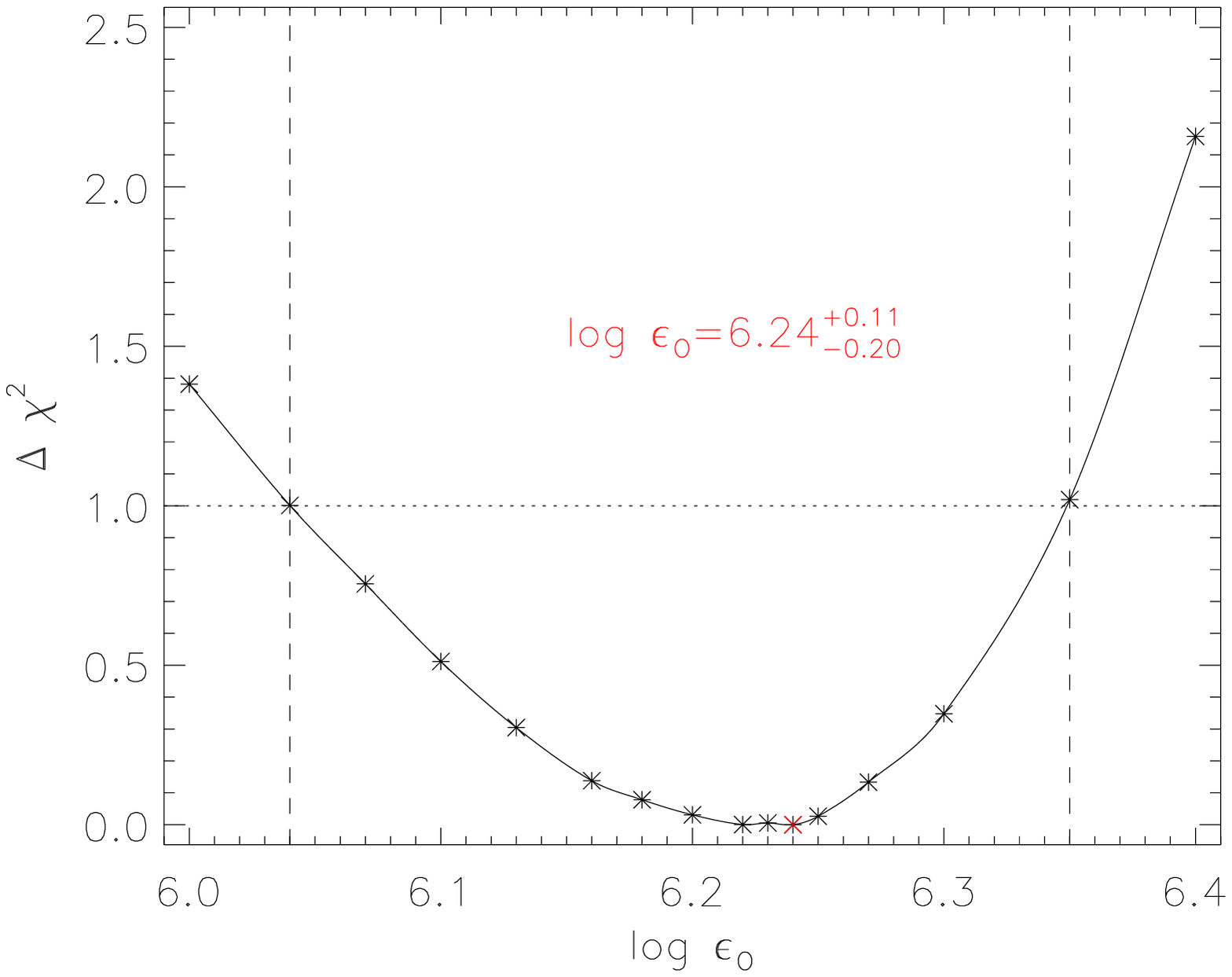}\\
  \caption{The spectrum synthesis of the \Oi\, IR triplet in the star
  CD-24~17504, for different oxygen abundances. A gaussian convolution
  profile with macroturbulence and instrumental broadening of 5 km~s$^{-1}$
  was adopted. {\it Left panel}: observed (thin) and computed (thick) line
  profiles. The latter are for our best estimate of the oxygen abundance (solid)
  and $\pm 1-\sigma$ (dashed), i.e. corresponding to $\log\epsilon_O=6.24^{+0.11}_{-0.20}$.
  {\it Right panel}: chi square variation (solid curve)
  and corresponding $1-\sigma$ uncertainties (dashed vertical lines).
  \label{fabf:fit}}
\end{figure*}

\end{flushleft}


\begin{flushleft}

\begin{table}
\caption{The measured equivalent widths W$_{\lambda}$ ({in m\AA})
for the \Ci\, and \Oi\, features detected in the 15 newly observed
halo stars used in this study, together with the S/N near the
features of interest. Only available online. \label{fabt:eqwidths}}

\end{table}

\end{flushleft}


The equivalent widths of all lines measured in this study are listed
in Table \ref{fabt:eqwidths}. The C\,{\sc i} lines near 9100\,\AA,
C\,{\sc i}~$\lambda 9405.7$, and the O\,{\sc i}~$\lambda \lambda
7772-7775$ triplet are all strong enough to be detected in our
high-resolution, high S/N spectra for most stars, even at the lowest
metallicities we explore. Akerman et al. (2004) used four
high-excitation \Ci\ features (at $9061.4$, $9078.3$, $9094.8$ and
$9111.8$~\AA) to determine the carbon abundance. In addition to
these lines, we also used C\,{\sc i}~$\lambda 9062.5$ and C\,{\sc
i}~$\lambda 9088.5$ available in the same spectral
region.\footnote{For C\,{\sc i}~$\lambda 9088.5$, we confirmed via
spectral synthesis that its blending with a weak \Fei\ line (at
$9088.3$~\AA) should not be neglected in the measurement of its
equivalent width and we have therefore taken it into account.}
Furthermore, we analysed all our spectra to detect the relatively
strong C\,{\sc i}~$\lambda 9405.7$ line (see
Fig.\,\ref{fabf:spectra}). This line is comparable in strength to
C\,{\sc i}~$\lambda 9094.8$ (the strongest in the $9100$~\AA\,
group), the difference in excitation potentials being almost
compensated by the difference in oscillator strengths (values are
taken from Wiese, Fuhr, \& Deters 1996, as retrieved using the NIST
Atomic Spectra Database version
3\footnote{Available online at:\\
http://physics.nist.gov/PhysRefData/ASD/index.html}).  It is thus
still detected at the lowest metallicities (\feh$\apprle-3$). We
found no indications of systematic differences in the carbon
abundance deduced from these additional C\,{\sc i} transitions
compared with those used by Akerman et al. (2004); by including
measurements of more C\,{\sc i} lines in our analysis, we improved
the accuracy of the determination of the carbon abundance,
especially at the lowest metallicities. Although only three  O\,{\sc
i} lines were analysed, they have the advantage of falling in a
spectral region which is free of contamination by telluric
absorption.

In the star \object{LP831-70}, which was observed with MIKE,
we could not detect with confidence any carbon or
oxygen line, despite the relatively high S/N achieved.
Taking for comparison the very
metal-poor star \object{CD-24~17504}, where the stronger
\Ci\  and \Oi\ lines are detected in the UVES spectra, we find that
\object{LP831-70} is cooler and less metal-poor. While the cooler
temperature goes in the direction of making the \Ci\, and \Oi\,
features weaker, the non-detections of
the \Ci\, and \Oi\, lines are still surprising, given
its higher metallicity (by about a factor
of two).

In general, the accuracy of the equivalent widths measurements
is about $\pm 1$~m\AA\ ($1 \sigma$), estimated empirically
by comparing the values measured on different spectra of
the same star, or from adjacent orders within the same spectrum
(\eg the \Oi\, triplet lines in our UVES
observations).
The equivalent widths listed in Table~\ref{fabt:eqwidths}
were employed to perform the LTE
abundance analysis through the use of the ``Uppsala package'' code
{\small EQWIDTH} and of {\small MARCS} model atmospheres, deriving,
for each star in the sample, the metallicity \feh\,
and the LTE C and O abundances.
The line-to-line scatter in the abundances we
obtain for carbon and oxygen is small (typically less than $\pm
0.05$~dex); at low \feh\, the main source of error is
the uncertainty in the equivalent widths of weak lines.
An uncertainty of $\pm 1$~m\AA\ in $W_{\lambda}$
translates to a random error
of  approximately 15\% in the abundances of C and O
in the most metal-poor stars.

\object{CD-24~17504} (alias G275-4) was a special case.
This star has the lowest metallicity
(\feh$=-3.21$) among those in the new sample
where we detect carbon lines.
As can be seen from Fig.~\ref{fabf:fit},
the \Oi\ lines are also very weak and almost at the
level of the noise---we detect with certainty only the bluest and
strongest transition of the \Oi\ triplet, $\lambda 7771.9$,
with an equivalent width $W_{\lambda} =  1.7$\,m\AA.
In order to put more
reliable limits on the oxygen abundance (and, thus, on [C/O]) in this
star, we carried out a spectral synthesis of the
\Oi\ triplet with the ``Uppsala package''
code BSYN for a range of values of the oxygen abundance.
We then performed a chi-square
minimization between the observed and synthesised spectra for
the wavelength region comprising the triplet lines, to find the best
estimate of the oxygen content of this star
(see right-hand panel of Fig.~\ref{fabf:fit}). The LTE result is
$\log\epsilon_O=6.24^{+0.11}_{-0.20}$, implying
[C/O]\,$= -0.16$.

In general, the LTE abundances we derive
seem to signal high values of up to
[C/O]\,$\sim -0.2$ at the very lowest metallicities.
In the following section, we apply non--LTE corrections computed
specifically for the stars in the sample, and for each of the lines we
have analysed, in order to assess the reality of the upturn in
\co\ at low values of \oh\ suggested by the LTE abundances.
By accounting for non--LTE effects in the formation
of the spectral lines of interest, we remove one of the main
sources of systematic error, although there are still
residual uncertainties in the
atomic (in particular, collision) data
used in the non--LTE analysis.

\subsection{\Ci\ and \Oi\ line formation: non--LTE abundance corrections
calculations}

The \Ci\ and \Oi\ spectral features used here are affected by
departures from LTE. Therefore, it is important to account for
non--LTE effects in order to derive reliable estimates of the
abundances of carbon and oxygen. Not only is it important to
understand how these effects work differentially between lines of
the same element, but in our case we obviously need to understand
how the different non--LTE corrections affect the behaviour of [C/O]
vs. [O/H] which is the main goal of this work.

Our recent investigations of C and O (Fabbian et al. 2006, 2008a)
have uncovered larger deviations from LTE for the \Oi\ lines than
for the \Ci\ lines. In Fabbian et al. (2006), we performed detailed
non--LTE calculations for the carbon lines which are employed also
in the present work. For the oxygen non--LTE corrections, in Fabbian
et al. (2006) we assumed the estimates by Akerman et al. (2004)  for
all but the five most metal-poor sample stars, for which we
re-estimated larger non--LTE abundance corrections. Such preliminary
investigation on the oxygen triplet subsequently led to the work
presented in Fabbian et al. (2008a), in which we additionally
introduced quantum-mechanical estimates for electron-impact
excitation. As discussed in Fabbian et al. (2008a), due to their
importance in coupling atomic levels of interest and thus forcing
large level overpopulation due to flow from radiatively pumped \Oi\
resonance lines, these new collisional data give much larger
non--LTE corrections for metal-poor turnoff stars, which we adopted
here. Unfortunately, no such quantum-mechanical calculations are as
yet available for carbon.\footnote{Collisions with electrons may
indeed be the most urgent outstanding problem to be addressed in the
formation of the \Ci\ lines, since including H collisions only
affects those features by $\apprle 0.1$~dex for typical stellar
parameters of interest. Given the similarity in the atomic
structure, one may expect that the \Ci\ IR lines are also
significantly affected by intersystem collisional coupling, and that
the quantum-mechanical electron-impact rates are larger than the
current estimates, as found in the case of oxygen. Our tests on
carbon indeed indicate a significant sensitivity to electron
collisions, including between singlet and triplet systems in the
atom, but especially to the ground state. However, an increase in
such rates tends to weaken the line, therefore {\it reducing} the
non--LTE effect on \Ci. We then expect that the high [C/O] values we
find in this work may be further increased, should more efficient
electron collisions be adopted for carbon too.}

Regarding the poorly known cross-sections for H collisions, while they
do not have a large impact on the \Ci\, non--LTE corrections found in
Fabbian et al. (2006) which we adopt here, and which amount to $\sim
-0.4$~dex in halo turnoff stars at \feh$\sim -3$, they are likely the
largest single remaining cause of uncertainty in the oxygen non--LTE
calculations. The results in Fabbian et al. (2008a) show that
neglecting collisions with neutral H atoms leads to the negative
abundance corrections becoming more severe (by up to $\sim 0.4$~dex).
Since oxygen is so sensitive to the choice of H collision efficiency,
in the absence of detailed quantum-mechanical calculations, one may
look for indirect evidence as to whether the classical recipe by
Drawin (1968, 1969) usually employed needs to be scaled by large
factors.  As discussed in Fabbian et al. (2008a), while some high
scaling factors have been suggested in the literature, we believe that
the Drawin recipe may indeed be a fairly accurate approximation for
oxygen, based on evidence in the Sun and from the fact that derived
[O/Fe] ratios would become unreasonably low and close to solar at low
metallicity if H collisions were neglected. The resulting oxygen non--LTE
corrections are then typically $\sim -0.5$~dex for metal-poor turnoff
stars. Non--LTE effects on the \Ci\, and \Oi\, lines thus work
differentially at low metallicity, giving higher [C/O] ratios by
$\sim +0.1$\,dex compared to LTE.

\section{Results}

\begin{flushleft}

\begin{table*}
\begin{center}
\begin{minipage}{\textwidth}
\caption{Derived abundances of oxygen and carbon for the
 sample of 43 halo stars. The resulting [C/H], [O/H] and [C/O] abundance
 ratios derived in this study are also given, both for LTE and
 non--LTE (using our adopted abundance corrections obtained with and
 without including collisions with H\,{\sc i} atoms, respectively).
 In the non--LTE case, S$_{\rm H}$ indicates the scaling factor
 regulating the efficiency of the collisions with neutral H atoms via the Drawin formula.
 The differential abundances ([C/O], [O/H], [Fe/H])
 with respect to the Sun were derived assuming$^{\rm 1}$
 $\log {\rm (C/H)_{\odot}} + 12 = 8.39$,
 $\log {\rm (O/H)_{\odot}} + 12 = 8.66$,
 and
 $\log {\rm (Fe/H)_{\odot}} + 12 = 7.45$.
 \label{fabt:sample}
 }
\begin{tiny}
\begin{tabular}{l|cc|ccc|ccc|ccc}
\hline \hline ID & $\log\epsilon_{\rm C}$ & $\log\epsilon_{\rm O}$
& [C/H] & [C/H] & [C/H] & [O/H] & [O/H] & [O/H] & [C/O] & [C/O] & [C/O]\\
& (LTE) & (LTE) & (LTE) & (S$_{\rm H}=0$) & (S$_{\rm H}=1$) & (LTE) & (S$_{\rm H}=0$) & (S$_{\rm H}=1$) & (LTE) & (S$_{\rm H}=0$) & (S$_{\rm H}=1$)\\
\hline
{\bf UVES (2001)} & & & & & & & & & & &\\
BD-13$^{\circ}$3442   & 6.14  & 6.89 & -2.25 & -2.62 & -2.56 & -1.77 & -2.39 & -2.11 & -0.48    & -0.23 & -0.45 \\
CD-30$^{\circ}$18140  & 6.73  & 7.58 & -1.66 & -1.93 & -1.84 & -1.08 & -1.27 & -1.22 & -0.58    & -0.66 & -0.62 \\
CD-35$^{\circ}$14849  & 6.38  & 7.05 & -2.01 & -2.30 & -2.22 & -1.61 & -1.99 & -1.81 & -0.40    & -0.31 & -0.41 \\
CD-42$^{\circ}$14278  & 6.54  & 7.28 & -1.85 & -2.07 & -1.97 & -1.38 & -1.55 & -1.47 & -0.47    & -0.52 & -0.50 \\
G011-044              & 6.58  & 7.37 & -1.81 & -2.03 & -1.94 & -1.29 & -1.47 & -1.40 & -0.52    & -0.56 & -0.54 \\
G013-009              & 6.48  & 7.19 & -1.91 & -2.25 & -2.17 & -1.47 & -1.83 & -1.73 & -0.44    & -0.42 & -0.44 \\
G018-039              & 7.23  & 8.04 & -1.16 & -1.40 & -1.29 & -0.62 & -0.84 & -0.77 & -0.54    & -0.56 & -0.52 \\
G020-008              & 6.54  & 7.30 & -1.85 & -2.11 & -2.01 & -1.36 & -1.63 & -1.51 & -0.49    & -0.48 & -0.50 \\
G024-003              & 6.65  & 7.56 & -1.74 & -1.97 & -1.87 & -1.10 & -1.30 & -1.24 & -0.64    & -0.67 & -0.63 \\
G029-023              & 6.82  & 7.71 & -1.57 & -1.84 & -1.75 & -0.95 & -1.16 & -1.11 & -0.62    & -0.68 & -0.64 \\
G053-041              & 7.01  & 7.78 & -1.38 & -1.66 & -1.53 & -0.88 & -1.09 & -1.03 & -0.50    & -0.57 & -0.50 \\
G064-012              & 5.67  & 6.45 & -2.72 & -3.17 & -3.09 & -2.21 & -3.10 & -2.71 & -0.51    & -0.07 & -0.38 \\
G064-037              & 5.71  & 6.42 & -2.68 & -3.13 & -3.05 & -2.24 & -3.12 & -2.70 & -0.44    & -0.01 & -0.35 \\
G066-030              & 6.91  & 7.90 & -1.48 & -1.76 & -1.66 & -0.76 & -1.03 & -0.96 & -0.72    & -0.73 & -0.70 \\
G126-062              & 6.94  & 7.86 & -1.45 & -1.72 & -1.62 & -0.80 & -1.04 & -0.98 & -0.65    & -0.68 & -0.64 \\
G186-026              & 6.15  & 6.75 & -2.24 & -2.56 & -2.49 & -1.91 & -2.47 & -2.20 & -0.33    & -0.09 & -0.29 \\
HD106038              & 7.38  & 8.05 & -1.01 & -1.28 & -1.14 & -0.61 & -0.82 & -0.75 & -0.40    & -0.46 & -0.39 \\
HD108177              & 6.90  & 7.75 & -1.49 & -1.73 & -1.63 & -0.91 & -1.11 & -1.04 & -0.58    & -0.62 & -0.59 \\
HD110621              & 7.02  & 7.89 & -1.37 & -1.65 & -1.54 & -0.77 & -0.99 & -0.94 & -0.60    & -0.66 & -0.60 \\
HD140283              & 6.27  & 6.99 & -2.12 & -2.40 & -2.32 & -1.67 & -1.91 & -1.81 & -0.45    & -0.49 & -0.51 \\
HD160617              & 6.71  & 7.39 & -1.68 & -1.97 & -1.87 & -1.27 & -1.47 & -1.42 & -0.41    & -0.50 & -0.45 \\
HD179626              & 7.54  & 8.38 & -0.85 & -1.11 & -0.98 & -0.28 & -0.51 & -0.45 & -0.57    & -0.60 & -0.53 \\
HD181743              & 6.76  & 7.58 & -1.63 & -1.84 & -1.74 & -1.08 & -1.24 & -1.18 & -0.55    & -0.60 & -0.56 \\
HD188031              & 6.84  & 7.69 & -1.55 & -1.81 & -1.72 & -0.97 & -1.18 & -1.12 & -0.58    & -0.63 & -0.60 \\
HD193901              & 7.41  & 8.17 & -0.98 & -1.19 & -1.06 & -0.49 & -0.65 & -0.58 & -0.49    & -0.54 & -0.48 \\
HD194598              & 7.44  & 8.15 & -0.95 & -1.22 & -1.11 & -0.51 & -0.75 & -0.68 & -0.44    & -0.47 & -0.43 \\
HD215801              & 6.32  & 7.22 & -2.07 & -2.37 & -2.29 & -1.44 & -1.72 & -1.61 & -0.63    & -0.65 & -0.68 \\
LP815-43              & 6.16  & 6.71$^{\rm a}$   & -2.23 & -2.63 & -2.56 & -1.95 & -2.70 & -2.36 & -0.28   & +0.07 & -0.20 \\
\hline
{\bf UVES (2004)} & & & & & & & & & & & \\
CD-24~17504$^{\rm b}$ & 5.81  & 6.24 & -2.58 & -2.91 & -2.84 & -2.42 & -3.25 & -2.87 & -0.16    & +0.34 & +0.03 \\
CD-71~1234            & 6.28  & 7.03 & -2.11 & -2.40 & -2.33 & -1.63 & -2.04 & -1.86 & -0.48    & -0.36 & -0.47 \\
CS 22943-0095         & 6.56  & 7.33 & -1.83 & -2.13 & -2.05 & -1.33 & -1.66 & -1.52 & -0.50    & -0.47 & -0.53 \\
G004-037              & 6.26  & 7.16 & -2.13 & -2.44 & -2.36 & -1.50 & -1.96 & -1.75 & -0.63    & -0.48 & -0.61 \\
G048-029$^{\rm c}$    & 6.01  & 6.81 & -2.38 & -2.70 & -2.64 & -1.85 & -2.48 & -2.18 & -0.53    & -0.22 & -0.46 \\
G059-027              & 6.81  & 7.60 & -1.58 & -1.84 & -1.74 & -1.06 & -1.24 & -1.18 & -0.52    & -0.60 & -0.56 \\
G126-052              & 6.42  & 7.15 & -1.97 & -2.26 & -2.18 & -1.51 & -1.83 & -1.69 & -0.46    & -0.43 & -0.49 \\
G166-054              & 6.03  & 6.91 & -2.36 & -2.72 & -2.64 & -1.75 & -2.33 & -2.05 & -0.61    & -0.39 & -0.59 \\
HD84937               & 6.53  & 7.27 & -1.86 & -2.15 & -2.07 & -1.39 & -1.64 & -1.56 & -0.47    & -0.51 & -0.51 \\
HD338529              & 6.46  & 7.24 & -1.93 & -2.25 & -2.17 & -1.42 & -1.77 & -1.63 & -0.51    & -0.48 & -0.54 \\
LP635-014             & 6.39  & 7.06 & -2.00 & -2.33 & -2.25 & -1.60 & -2.03 & -1.85 & -0.40    & -0.30 & -0.40 \\
LP651-004             & 6.08  & 7.04 & -2.31 & -2.99 & -2.58 & -1.62 & -2.21 & -1.93 & -0.69    & -0.44 & -0.65 \\
\hline
{\bf MIKE (2003)} & & & & & & & & & & &\\
G041-041             & 5.96   & 6.74 & -2.43 & -2.85 & -2.76 & -1.92 & -2.54 & -2.25 & -0.51    & -0.31 & -0.51 \\
G048-029$^{\rm c}$   & 6.11   & 6.80 & -2.28 & -2.66 & -2.59 & -1.86 & -2.50 & -2.20 & -0.42    & -0.16 & -0.39 \\
G084-029             & 6.06   & 6.87 & -2.33 & -2.72 & -2.63 & -1.79 & -2.33 & -2.08 & -0.54    & -0.39 & -0.55 \\
LP831-070$^{\rm b}$  & $<5.77$ & $<6.53$ & $<-2.62$ & $<-3.01$ & $<-2.93$ & $<-2.13$ & $<-2.95$ & $<-2.54$ & - & - & - \\
\hline
 & & & & & & & & & & & \\

\end{tabular}
\end{tiny}

$^{\rm 1}$ Asplund, Grevesse, \& Sauval 2005.

\vspace{0.1cm}

$^{\rm a}$ As discussed by Fabbian et al. (2006), the
oxygen determination by Nissen et al. (2002) for this stars
is more reliable than that
derived from the UVES 2001 spectra and was therefore adopted here.

\vspace{0.1cm}

$^{\rm b}$ Since the \Oi\, lines are barely visible in these two
stars, spectral synthesis and chi square minimization (see text) of
the profiles of the \Oi\, triplet lines were used in order to
constrain their oxygen content.

\vspace{0.1cm}

$^{\rm c}$ This star was observed in both UVES (2004) and MIKE
(2003) runs. The C and O abundances derived in the two cases using
the respective atmospheric parameters for the two sets of spectra
are shown here, while their mean was adopted in
Figs.\,\ref{fabf:cfe}, \ref{fabf:ofe}, \ref{fabf:cooh} and
\ref{fabf:feooh}.

\end{minipage}
\end{center}
\end{table*}

\end{flushleft}

\begin{flushleft}

\begin{figure}
\begin{center}
    \includegraphics[width=8.9cm]{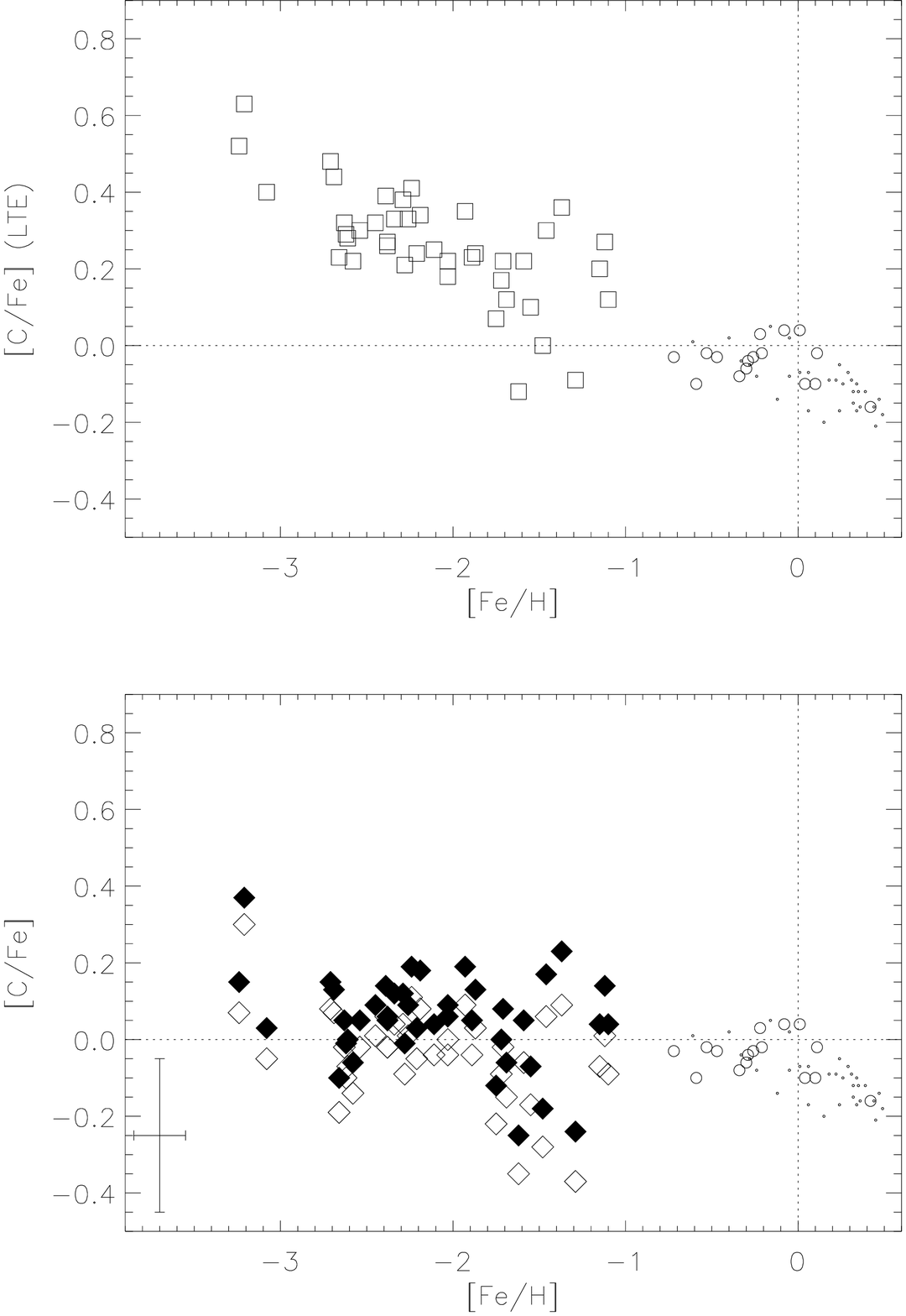}
    \caption{Runs of [C/Fe] versus \feh\ for our sample stars,
    respectively in LTE (upper panel) and non--LTE (lower panel: filled
    diamonds\,=\,H collisions 'a la Drawin'; empty diamonds\,=\,no H collisions).
    The results by Bensby \& Feltzing (2006), obtained from
    [C\,{\sc i}] and [O\,{\sc i}] lines free from non--LTE effects, are
    additionally shown for thin (small dots) and thick (open circles)
    disc stars. The dotted horizontal and vertical lines indicate the solar
    values.
    \label{fabf:cfe}
    }
\end{center}
\end{figure}

\end{flushleft}

\begin{flushleft}

\begin{figure}
\begin{center}
    \includegraphics[width=8.9cm]{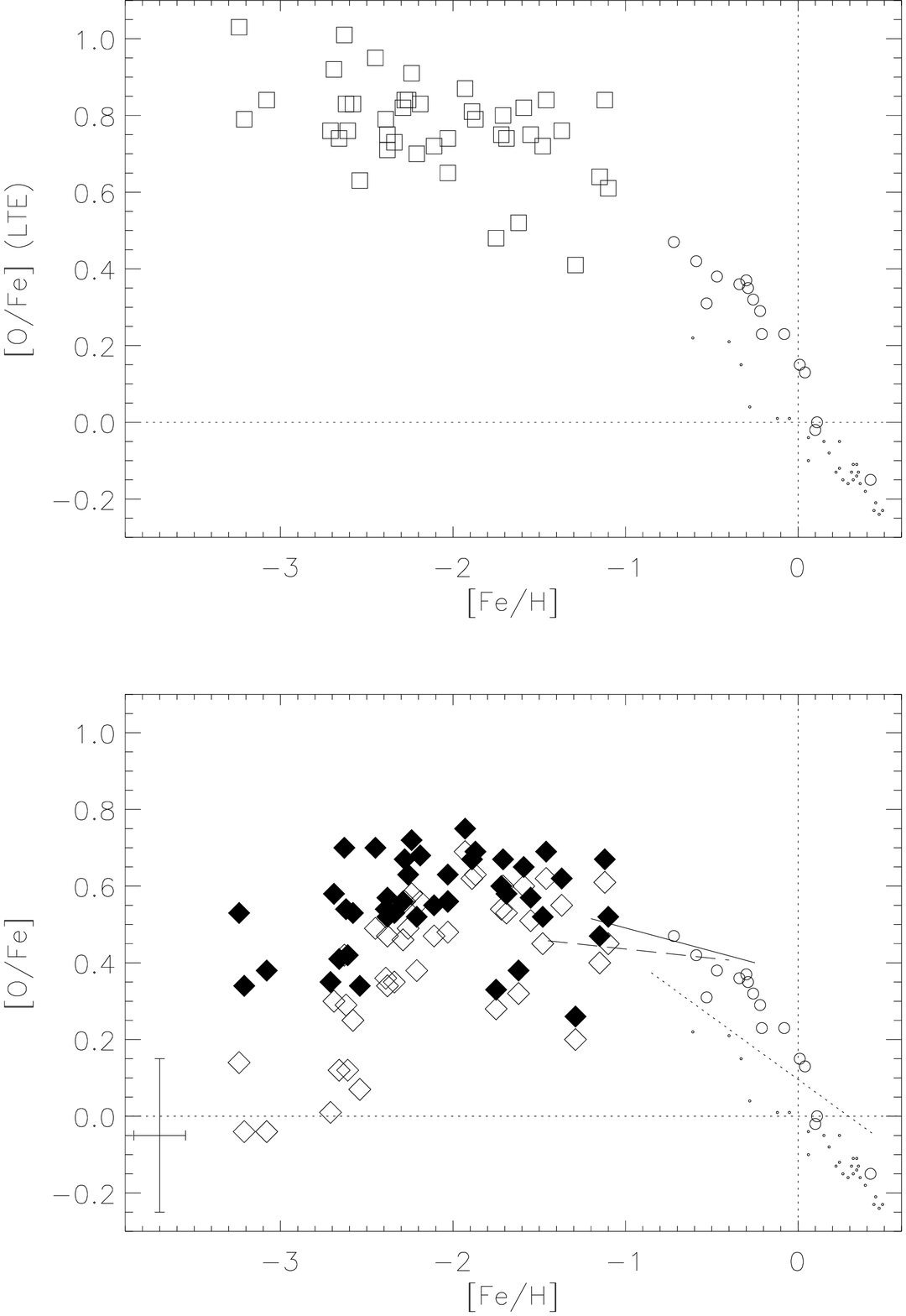}
    \caption{Runs of [O/Fe] versus \feh\ in LTE (upper panel) and
    non--LTE (lower panels), respectively. Symbols are as in
    Fig.\,\ref{fabf:cfe}. Lines in the bottom panel show the fits
    derived by Ram{\'\i}rez et al. (2006) to their [O/Fe] data, separately
    for thin disc (dotted line), thick disc
    (solid line) and halo (dashed line) stars.
     \label{fabf:ofe}
     }
\end{center}
\end{figure}

\end{flushleft}

\begin{flushleft}

\begin{figure}
    \begin{center}
    \includegraphics[width=8.9cm]{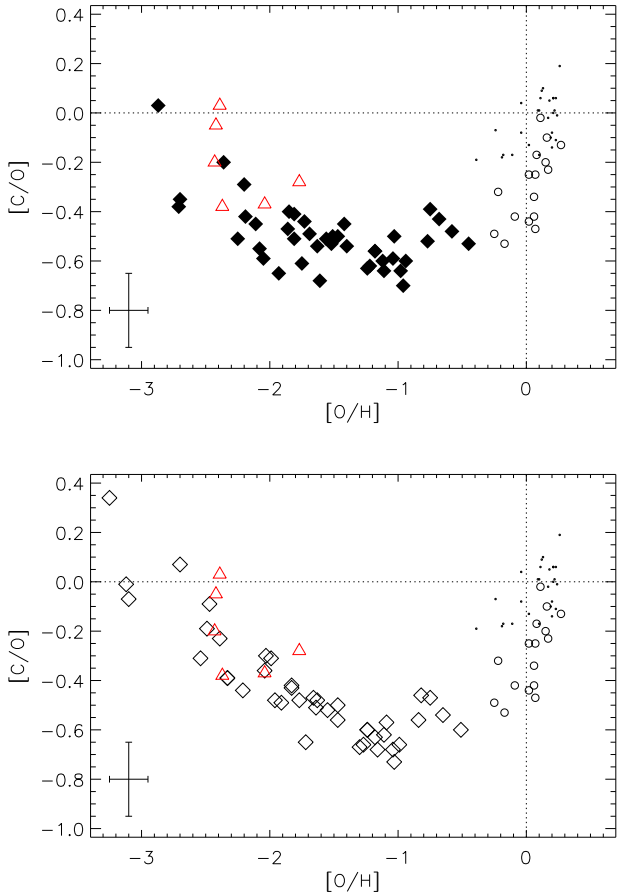}
    \caption{Final estimates of [C/O] versus [O/H] for our sample stars
     using, for both carbon and oxygen, S$_{\rm H}=1$
    (upper panel), or neglecting H collisions altogether (lower panel).
    The meaning of the symbols for the different Galactic stellar components
    is as in Fig.~\ref{fabf:cfe}, lower panel.
    The data indicated by triangles are the values measured
    in metal-poor damped Lyman alpha systems at high redshifts by Pettini
    et al. (2008).
    \label{fabf:cooh}}
\end{center}
\end{figure}

\end{flushleft}

\begin{flushleft}

\begin{figure}
    \begin{center}
    \includegraphics[width=8.9cm]{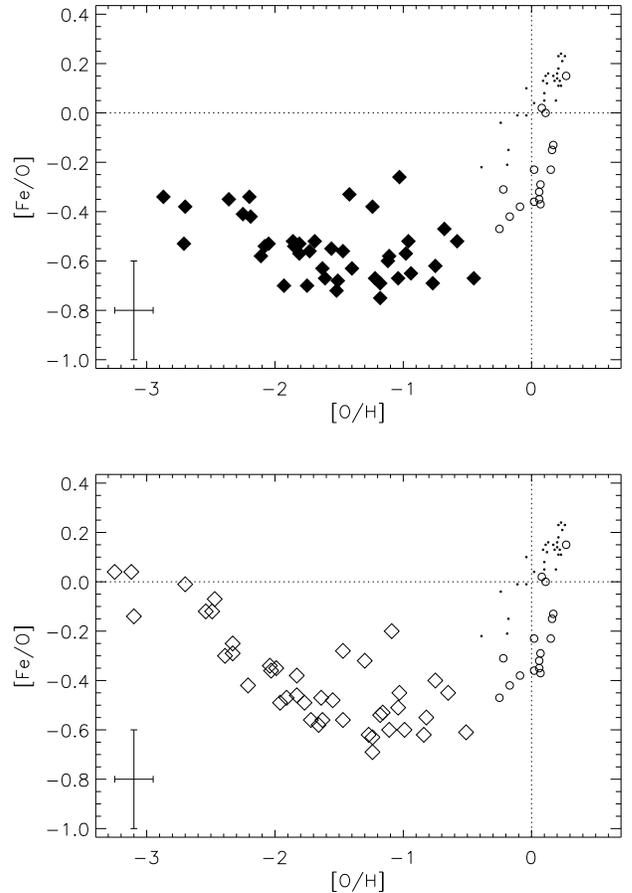}\\
    \caption{
    Final estimates of [Fe/O] versus [O/H] for our sample stars,
    using for oxygen either S$_{\rm H}=1$ (upper panel) or
    neglecting H collisions altogether (lower panel).
    The meaning of the symbols for the different Galactic stellar components
    is as in Fig.~\ref{fabf:cfe}, lower panel.
    \label{fabf:feooh}}
\end{center}
\end{figure}

\end{flushleft}

Our final abundance results are given in Table \ref{fabt:sample}.
Figures~\ref{fabf:cfe} and \ref{fabf:ofe} show the corresponding
trends of [C/Fe] and [O/Fe] with [Fe/H], while Figs. \ref{fabf:cooh}
and \ref{fabf:feooh} show the behaviour of [C/O] and [Fe/O] with
[O/H].

The typical errors associated with our abundance estimates for the
most metal-poor stars in the sample (where the uncertainties are
largest due to fewer and weaker \Ci\ and \Oi\ lines being measured)
are indicated by the error bars in the lower left-hand corner of
Figs.~\ref{fabf:cfe}, \ref{fabf:ofe}, \ref{fabf:cooh} and
\ref{fabf:feooh}. They correspond to the statistical error
introduced by the $1-\sigma$ errors in \Teff\, ($\pm 100$~K),
\logg\, ($\pm 0.15$~dex), microturbulence ($\pm 0.3$~km/s) and
equivalent width. In particular, the latter measurement errors
dominate for the most metal-poor stars. The uncertainty on \Teff\,
is significant too, in the case of C/Fe and O/Fe, but cancels out in
the case of C/O.

Referring to Fig.~\ref{fabf:cfe}, it is interesting to note that the
introduction of non--LTE corrections in the analysis of the C\,{\sc
i} lines completely erases the LTE trend of increasing [C/Fe] with
decreasing [Fe/H]. The essentially flat behaviour of [C/Fe] at
near-solar values over three orders of magnitude in [Fe/H] seen in
the lower panel of Fig.~\ref{fabf:cfe} would indicate that C and Fe
production in the Galaxy has proceeded on similar timescales and
thus presumably from similar sources.

Regarding the widely debated behaviour of [O/Fe] at low metallicity,
our non--LTE corrected abundances (Fig.~\ref{fabf:ofe}) seem to back
the idea that the real trend may be essentially flat, with
abundances derived from the \Oi\, $7772-7775$~\AA\, triplet lying
much closer than often reported to those usually derived in the
literature from [\Oi] and infrared OH lines. When accounting for
full 3D non--LTE effects on {\it all} the affected lines becomes
possible, it is quite plausible that agreement may be finally
reached. At this stage, we can only comment on the fact that the
application of non--LTE corrections without taking into account H
collisions seems to overcorrect the LTE abundances and yield values
of [O/Fe] which are almost certainly too low, being close to solar
at [Fe/H]\,$\apprle - 2.5$ (see Fig.~\ref{fabf:ofe}). We consider it
more likely that H collisions are fairly efficient in the case of
oxygen, making the relevant non--LTE corrections less severe and
only slightly larger than for carbon. In addition, based on the
results of solar observations by Allende Prieto et al. (2004), we
can safely rule out that LTE (S$_{\rm H} \gg 1$) applies to the
\Oi\, $7772-7775$~\AA\, triplet. It remains of high priority to
carry out full non--LTE calculations with hydrodynamical model
atmospheres for oxygen at low metallicity, hopefully including
future quantum-mechanical calculations of inelastic H collisions.

Turning now to the [C/O] ratio, non--LTE results for our sample are
shown in Fig. \ref{fabf:cooh}, for two different choices of S$_{\rm
H}$, together with those for  higher-metallicity disc stars obtained
by Bensby \& Feltzing (2006) from forbidden \Ci\ and \Oi\ lines. The
new data strengthen the suggestion by Akerman et al. (2004) that the
decrease of [C/O] between solar and intermediately-low metallicities
(i.e., the metallicity range of thin and thick disc), reaching a
minimum of \co$\sim -0.7$ at \oh$\sim -1.0$ in our data, turns into
an \emph{increase} in halo stars of even lower metallicities. The
adoption of non--LTE corrections tends to move the data points for
the metal-poor halo stars to: (a) much lower values of [O/H] due to
large oxygen non--LTE corrections, and (b) to higher [C/O] values
because the negative non--LTE corrections are $0.1-0.4$\,dex more
severe for oxygen than for carbon, depending on the choice of H
collision efficiency and on the particular stellar parameters. This
``stretches'' the rising trend seen in LTE towards lower
metallicities, while at the same time raising [C/O] to
close-to-solar values. The rise has a slope of $\sim -0.3$ in the
[C/O] vs. [O/H] plane.

We have included in Fig. \ref{fabf:cooh}
the [C/O] measurements in metal-poor DLAs
by Pettini et al. (2008) which seem to match well the
values deduced here in halo stars of similar [O/H].
The good agreement in the [C/O] ratios measured
in different astrophysical
environments and at different epochs
strengthens the interpretation
that carbon was somehow overproduced in early stages of
galactic chemical evolution.
On the other hand, the halo giants considered by
Spite et al. (2005) (we refer here to the ``unmixed''
objects from their sample) appear to
have generally lower [C/O] values than those derived here,
particularly in view of the large 3D effects which are likely
to apply, in giants, to the
CH features employed in their analysis (Collet et al. 2007).
Such corrections would decrease their [C/O]
determinations by several tenths of a dex.

Further investigations targeting stars at different evolutionary
stages are bound to help shed light on this issue. The star with the
highest \co\ in our sample is \object{CD-24~17504},  for which
we deduce \feh$\,=-3.21$, and
\co\,$=+0.03$ or $+0.34$, depending on whether the
effects of H collisions are included or not.  Richard, Michaud \&
Richer (2002) specifically discuss this star as peculiar in relation
to atomic diffusion effects, arguing that these could have modified
its relative metal abundances, so that caution should be exercised
when interpreting its derived abundance ratios in terms of
nucleosynthesis. In particular, their Figs. 5 and 11 show that
differential effects on C and O may be important, mainly due to larger
oxygen surface abundance decrease.

The other two stars with metallicities
below \feh$\,=-3$ (\object{G64-12} and \object{G64-37} from
the Akerman et al. 2004 sample)
have lower carbon enhancements
with [C/O]\, $\sim -0.35$ or $\sim 0$, again depending
on whether H collisions are included or not.
In the case of \object{LP831-70}
([Fe/H]\,$=-2.94$), our spectrum is too noisy for
positive detections of the weak C\,{\sc i} and O\,{\sc i}
lines. Our conservative upper limits (Tables \ref{fabt:eqwidths}
and \ref{fabt:sample}) suggest that this star has
lower C and O abundances than other stars of similar metallicity.

Finally, we show the trend of [Fe/O] with
[O/H]  in Fig.\,\ref{fabf:feooh}. It is interesting to compare this diagram
with Fig.\,\ref{fabf:cooh}. If carbon and iron were produced mainly by
stars with similar evolutionary timescales,
one would expect the figures to look very
similar, as indeed they do. \\


\section{Galactic chemical evolution of carbon and oxygen}
\label{fabs:GCE}

Stars born at early galactic epochs and living through to the
present time bear the signatures in their chemical composition (in
particular in the relative abundances of different elements as a
function of metal content) of events occurring during the history
of the Milky Way, providing information on its different formation
events (Freeman \& Bland-Hawthorn 2002) and on the nucleosynthetic
channels that build the various elements in the interiors of stars.

While it is known that carbon is produced during helium burning in
stars of all masses, the sites of major contribution to the C
enrichment in our Galaxy and the dependence of yields on stellar
mass, which is used as input to Galactic Chemical Evolution (GCE)
models describing how the abundances of various elements vary in
time, are not yet agreed upon. The roughly flat [C/Fe] trend we see
down to low metallicities seems to suggest that the sources of
carbon and iron have similar timescales, at least until the
metallicity becomes extremely low. At variance with the case of
oxygen, stars of different masses and thus operating on different
timescales have contributed to the build-up of carbon during the
Galaxy's history. Standard GCE models predict that the time lag
between the prompt ISM enrichment of oxygen due to short-lived
massive stars exploding as type II supernovae, and the delayed
released of carbon, should cause C/O to constantly decrease with
decreasing metallicity. In particular, Carigi et al. (2005) have
suggested that yields from massive stars are the overwhelming source
of carbon at early stages, while later on, at the end of their
slower evolution, low- and intermediate-mass stars would be able to
contribute carbon ejecta into the ISM in a comparable amount. In
their models, a combination of sources differing in mass and thus
contributing at different times is required to match the observed
trends. In contrast, Gavil\'{a}n, Buell \& Moll\'{a} (2005) have
argued that low- and intermediate-mass stars alone may account for
the carbon evolution.

Given the very low metal-content of most stars in our sample, with
oxygen abundances as low as $10^{-3}$ of the solar value, these objects
are presumably  associated with very early star formation in our Galaxy, before
the end of the halo build-up. Thus, by being the only survivors to
the present, they provide a window on nucleosynthetic processes
taking place in the stars that have existed at such early times.
A straightforward interpretation of the high C/O values we find
at low metallicities is that the first episodes of star formation in the
Galaxy provided a
source of high C abundance, perhaps thanks to a primordial
generation of massive, zero-metallicity stars. According to current
theoretical models describing the yields of these hypothetical
objects, it is plausible that they could have indeed contributed
large C yields (e.g. Chieffi \& Limongi 2002, 2004).

Akerman et al. (2004) constructed GCE models in order to interpret
their tentative discovery of a [C/O] rise at low metallicity. By
adopting the Population III yields of Chieffi \& Limongi (2002), they
could reproduce the observed behaviour, in particular when using a
top-heavy IMF. This would imply, as assumed in the derivation of
those yields, that the nucleosynthetic channel
$^{12}$C($\alpha$,$\gamma$)$^{16}$O proceeds at a lower rate in
such primordial objects.

Chiappini et al. (2006) on the other hand argue that the [C/O]
upturn can be explained through fast stellar rotation at very low
metallicities, so that due to lower average core temperature, the
conversion of C into O would be less efficient. However, they also
make it clear that it is not granted that the high C/O values should
necessarily imply the signature of massive Pop. III stars, since
their own results can be achieved without including zero-metallicity
yields.

Carigi et al. (2005) successfully fitted the observed radial
gradients of C/O and O/H in the Milky Way with models
which use a steep IMF and in which the relative proportions
of carbon released into the interstellar medium by
massive stars on the one hand, and  low- and intermediate-mass stars
on the other, vary with time and galactocentric distance. Their models
are  in reasonably good agreement with the observed trends of
the ratios [C/O], [C/Fe] and [O/Fe] reported here.

Nissen et al. (2007) reported evidence for an increase
in the abundance of Zn relative to Fe at the lowest metallicities,
with [Zn/Fe]\,$\simeq +0.5$ at [Fe/H]\,$\apprle -3$.
Traditional yields of Type II SNe (Nomoto et al. 1997) cannot reach
the high observed [Zn/Fe] values, which instead seem to require
either the ejecta of Population III hypernovae,
or high Zn production from core-collapse, very
massive ($M \sim 500{-}1000\,M_{\odot}$) stars
(Ohkubo et al. 2006).

Kobayashi et al. (2006) calculated yields for a wide range of
metallicities ($Z =0 - Z_{\odot}$) and explosion energies,
including hypernovae. Their predicted \cfe--\feh\, and
\ofe--\feh\, relations seem to match our derived abundances for halo
stars (assuming S$_{\rm H}=0$ and $1$ for carbon and oxygen
respectively) fairly closely. In particular for carbon, the agreement
seems to indicate that enrichment from stellar winds and the
contribution of low- and intermediate-mass stars are not important for
this element at low metallicity, since those are not included in the
calculations by Kobayashi et al. (2006). Those authors pointed out
that in order to explain simultaneously the high C abundances observed
in extremely metal-poor stars and the supersolar Zn/Fe ratios at low
metallicities, other enrichment sources (e.g. a few Pop. III supernova
explosions in the very early inhomogeneous intergalactic medium, or
external enrichment from a binary companion) are needed. The final C
yields may in fact turn out to be even higher than predicted by
Kobayashi and collaborators, if winds of massive stars at very low
metallicity contribute significant additional amounts of carbon.

Finally, Smiljanic et al. (2008) very recently discussed possible signatures
of hypernova nucleosynthesis in \object{HD~106038}, one of the stars
in our sample, for which we derive ${\rm [Fe/H]} = -1.37$.
Even though
their hypothesis is mainly based on very large beryllium
enhancement, they also discuss available literature values for other
elements. For carbon and oxygen in this star, we find
\cfe\,= +0.23/+0.09 and
\ofe\,= +0.62/+0.55, depending on
whether H collisions are included or not.
The values of \ofe\ we deduce are in
good agreement
with the non--LTE determination by Mel\'{e}ndez et al.
(2006) of \ofe\,$=+0.56 \pm 0.10$.

\section{Conclusions}

We have presented non--LTE corrected element abundances in a sample
of 43 metal-poor halo stars, and carried out the most extensive
study to date of the relative abundances of carbon and oxygen as a
function of metallicity. Updated estimates of stellar parameters for
stars with \feh$< -1$ in the sample of Akerman et al. (2004) were
derived consistently with the rest of our sample stars following
Nissen et al. (2007). Akerman et al. (2004) estimated that,  to a
first approximation, non--LTE and 3D effects on the formation of the
C\,{\sc i} and O\,{\sc i} lines they analysed are of similar
magnitude so that their neglect should not lead to a systematic bias
in the values of [C/O] deduced. However, they also warned that a
quantitative assessment of non--LTE effects was required to confirm
the reality of their tentative findings. Here we have included
non--LTE corrections for both C and O and considered additional
C\,{\sc i} lines; taken together these two aspects of the present
work represent significant improvements over earlier published
studies by reducing both systematic and statistical uncertainties in
the determinations of the C and O abundances. The sample of stars
analysed by Akerman et al. (2004) was subsequently used by
Mel\'{e}ndez et al. (2006) in their study of oxygen at low
metallicity. However, while Mel\'{e}ndez and collaborators applied
their non--LTE corrections directly to published values, we have
re-derived the abundances to complement our sample of newly observed
stars in a consistent fashion. Our main findings are as follows.

\begin{enumerate}

\item After accounting for large negative non--LTE effects on the \Ci\,
and \Oi\, lines employed here, our resulting abundances reinforce
and place on stronger footing the case for a trend of rising
[C/O] with decreasing [O/H] in Galactic halo stars
first discovered by Akerman et al. (2004).
Our improved analysis finds the strongest evidence so far
for an increase in [C/O] at \oh$\la -2$. This seems to
add to the suggestion that high [C/O] values are commonplace at low
metallicities, as recently argued by Pettini et al. (2008) in their
investigation of high-redshift absorption systems.

\item The exact magnitudes of the corrections to the abundances
of C and O from non--LTE effects are still uncertain because of
poorly known H collisions, especially for oxygen. The non--LTE
effects on the C\,{\sc i} and O\,{\sc i} lines we have analysed are
both negative and thus affect less the determination of the [C/O]
ratio than the individual abundances of the two elements. However,
the effects do not cancel out completely, with the corrections for
the O\,{\sc i} lines being larger than those for the C\,{\sc i}
lines in the low-metallicity regime we are most interested in.
Consequently, consideration of non--LTE effects leads to residual
\emph{positive} corrections to the values of [C/O] compared to LTE
analyses which tend to {\it underestimate} [C/O].

\item While detailed 3D corrections are not available,
we do not expect them to change our
results for [C/O] much, because determination
of this ratio from high excitation
\Ci\, and \Oi\, lines is virtually insensitive to temperature
changes in the atmospheric model structure, and because 3D abundance
corrections for these lines are expected to be small, of the same
sign, and of similar magnitude, and should therefore cancel out to a large extent
in the derivation of [C/O]. There may of course be some complicated
differential effects due to the coupling between non--LTE and 3D.
Full 3D non--LTE computations at very low metallicity using the new
generation of hydrodynamical model atmospheres and addressing this
remaining uncertainty are beyond the scope of this paper. However,
given the important role in galactic chemical evolution, targeting
these elements remains a priority for future investigations.

\item We find that the [C/O] ratio reduces by a factor of $3-4$ when
[O/H] decreases from solar to $\sim 1/10$ solar (as already shown
e.g. by Gustafsson et al. 1999 and as predicted by GCE models due
mainly to metallicity-dependent theoretical carbon yields from
winds/mass loss in metal-rich massive stars). At still lower
metallicities, [C/O] tends to increase---with a slope of $\sim
-0.3$ in the [C/O] vs. [O/H] plane---reaching near-solar values
again at  [O/H]\,$\sim -3$.

\end{enumerate}

Our results likely signal non-standard carbon and/or oxygen
nucleosynthesis, offering potentially crucial clues on the early
history of the Milky Way. Massive Population III stars may be the
sources responsible for such high \co\ ratios, with large carbon
yields from a generation of (so far elusive) primordial, massive,
metal-free objects. Alternatively, Population II stars in which high
production of C may have been aided by strong mass loss induced by
fast rotation could also be called upon to explain our results.
Models of fast-rotating, massive stars (so called ``spinstars'')
developed recently (Meynet et al. 2007) may provide an explanation
to our observations. Implications in terms of early generations of
stars at low metallicity including more fast-rotators than in the
present Universe will need to be confirmed by future observational
and theoretical studies.

\bigskip
\begin{acknowledgements}
DF acknowledges the hospitality of
the Department of Physics and Astronomy of the University
of Aarhus, Denmark. We are grateful to Jorge Mel\'{e}ndez
for fruitful discussions on the [O/Fe] ratio and for advice
on the IRFM temperature scale, and to Anna Frebel for her help in estimating
colour excesses and for sharing updated C and O abundance estimates
in HE~1327~2326 in advance of publication.
We also thank Kurt Adelberger for the
observations with the Magellan telescope.
This work has been partly funded by the Australian
Research Council (grants DP0342613 and DP0558836).
\end{acknowledgements}

\end{document}